\documentclass[aps]
{revtex4}

\usepackage[utf8]{inputenc} 
\usepackage{amsmath}
\usepackage{amssymb}
\usepackage{mathrsfs}
\usepackage{bm}
\usepackage{color}

\usepackage{graphicx}
\DeclareGraphicsExtensions{.pdf,.jpg}

\newcommand{\ket}[1]{| #1 \rangle}
\newcommand{\bra}[1]{\langle #1 |}

\newcommand{\tr}{\text{Tr}}

\begin{document}
\author{Bruno Mera$^{1,2,4}$, Chrysoula Vlachou$^{3,4}$, Nikola Paunkovi\'c$^{3,4}$, V\'{\i}tor R. Vieira$^{1,2}$ and Oscar Viyuela$^{5,6}$}
\affiliation{$^1$ CeFEMA, Instituto Superior
T\'ecnico, Universidade de Lisboa, Av. Rovisco Pais, 1049-001 Lisboa, Portugal}
\affiliation{$^2$  Departamento de F\'{\i}sica, Instituto Superior
T\'ecnico, Universidade de Lisboa, Av. Rovisco Pais, 1049-001 Lisboa, Portugal}
 \affiliation{$^3$ Departamento de Matemática, Instituto Superior Técnico, Universidade de Lisboa, Av. Rovisco Pais, 1049-001 Lisboa, Portugal}
\affiliation{ $^4$ Instituto de Telecomunica\c{c}\~oes, 1049-001 Lisbon, Portugal}
\affiliation{ $^5$ Department  of  Physics,  Harvard  University,  Cambridge,  MA  02318,  USA}
\affiliation{ $^6$ Department  of  Physics,  Massachusetts  Institute  of  Technology,  Cambridge,  MA  02139,  USA}
\title{Dynamical phase transitions at finite temperature from fidelity and interferometric Loschmidt echo induced metrics}
\begin{abstract}
We study finite-temperature Dynamical Quantum Phase Transitions (DQPTs) by means of the fidelity and the interferometric Loschmidt Echo (LE) induced metrics. We analyse the associated dynamical susceptibilities (Riemannian metrics), and derive analytic expressions for the case of two-band Hamiltonians. At zero temperature the two quantities are identical, nevertheless, at finite temperatures they behave very differently. Using the fidelity LE, the zero temperature DQPTs are gradually washed away with temperature, while the interferometric counterpart exhibits finite-temperature Phase Transitions (PTs). We analyse the physical differences between the two finite-temperature LE generalisations, and argue that, while the interferometric one is more sensitive and can therefore provide more information when applied to genuine quantum (microscopic) systems, when analysing many-body macroscopic systems, the fidelity-based counterpart is a more suitable quantity to study. Finally, we apply the previous results to two representative models of topological insulators in 1D and 2D.
\end{abstract}
\maketitle

\section{Introduction}
\label{sec:Intro}

Equilibrium Phase Transitions (PTs) are characterised by a non-analytic behaviour of relevant thermodynamic observables with respect to the change of temperature. Quantum Phase Transitions (QPTs)~\cite{sac:07}, traditionally described by Landau theory~\cite{lan:37}, occur when we adiabatically change a physical parameter of the system at zero temperature, i.e., the transition is driven by purely quantum fluctuations. A particularly interesting case arises when one studies QPTs in the context of topological phases of matter~\cite{wen:90, kos:tho:72, kos:tho:73, hal:83, hal:88}. They are quite different from the standard QPTs, since they do not involve symmetry breaking and are characterised by global order parameters~\cite{tknn:82}. These novel phases of matter, which include topological insulators and superconductors~\cite{has:kan:10, qi:zha:11, ber:hug:13}, have potentially many applications in emerging fields such as spintronics, photonics or quantum computing. Although many of their remarkable properties have traditionally been studied at zero temperature, there has been a great effort to generalise these phases from pure to mixed states, and finite temperatures~\cite{bar:waw:alt:flei:die:17, grus:17, mer:vla:pau:vie:17,mer:vla:pau:vie:17:qw, sjo:15, viy:riv:mar:15, nie:hub:14, viy:riv:del:14,viy:riv:del:2d:14, bar:bar:krau:rico:ima:zol:die:13,riv:viy:del:13, viy:riv:del:12,viy:riv:gas:wal:fil:del:18}.

The real time evolution of closed quantum systems out of equilibrium has some surprising similarities with thermal phase transitions, as noticed by Heyl, Polkovnikov and Kehrein~\cite{hey:pol:keh:13}. They coined the term Dynamical Quantum Phase Transitions (DQPTs) to describe the non-analytic behaviour of certain dynamical observables after a sudden quench in one of the parameters of the Hamiltonian. Since then, the study of DQPTs became an active field of research and a lot of progress has been achieved in comparing and connecting them to the equilibrium PTs~\cite{kar:sch:13,and:sir:14,vaj:dor:14,hey:15,kar:sch:17,hal:zau:17,zau:hal:17,hom:abe:zau:hal:17}. Along those lines, there exist several studies of DQPTs for systems featuring non-trivial topological properties~\cite{vaj:dor:15,sch:keh:15,bud:hey:16,hua:bal:16,jaf:joh:17,sed:flei:sir:17}. DQPTs have been experimentally observed in systems of trapped ions~\cite{jur:etal:17} and cold atoms in optical lattices~\cite{fla:etal:16}. The figure of merit in the study of DQPTs is the Loschmidt Echo (LE) and its derivatives, which have been extensively used in the analysis of quantum criticality~\cite{qua:son:liu:zan:sun:06,zan:pau:06,zan:qua:wan:sun:07,pol:muk:gre:moo:10,jac:ven:zan:11} and quantum quenches~\cite{ven:zan:10}. At finite temperature, generalisations of the zero temperature LE were proposed, based on the mixed-state Uhlmann fidelity~\cite{ven:tac:san:san:11,jac:ven:zan:11}, and the interferometric mixed-state geometric phase~\cite{hey:bud:17, bah:ban:dut:17}.
 For an alternative approach to finite-temperature DPTs, see~\cite{lan:fra:hal:17,hal:zau:mcc:veg:sch:kas:17}. 
 Fidelity is a measure of state distinguishability, which has been employed numerous times in the study of PTs~\cite{zan:pau:06,pau:sac:nog:vie:dug,zan:ven:gio:07,aba:ham:zan:08,zha:zho:09}, while the interferometric mixed-state geometric phase was introduced in~\cite{sjo:pat:eke:ana:eri:oi:ved:00}. The two quantities are in general different and it comes as no surprise that they give different predictions for the finite temperature behaviour~\cite{bah:ban:dut:17}: the fidelity LE does not show DPTs at finite temperatures, while the interferometric LE indicates their persistence at finite temperature. Thus, it remains unclear what the fate of DQPT at finite temperature truly is, and which of the two opposite predictions better captures the many-body nature of these PTs.

In this paper, we discuss the existence of finite temperature Dynamical Phase Transitions (DPTs) for the broad class of two-band Hamiltonians in terms of both the {\em fidelity} and the {\em interferometric LEs}. We derive analytic expressions for the metrics (susceptibilities) induced by the fidelity and the interferometric LEs, respectively, showing explicitly that the two approaches give different behaviours: the fidelity susceptibility shows a gradual disappearance of DPTs as the temperature increases, while the interferometric susceptibility indicates their persistence at finite temperature (consistent with recent studies on interferometric LE~\cite{hey:bud:17, bah:ban:dut:17}). 
We analyze the reasons for such different behaviours. The fidelity LE quantifies state distinguishability in terms of measurements of physical properties, inducing a metric over the space of quantum states, while the interferometric LE quantifies the effects of quantum channels acting upon a state, inducing a pullback metric over the space of unitaries. Thus, we argue that the fidelity LE and its associated dynamical susceptibility are more suitable for the study of many-body systems, while the more sensitive interferometric counterparts are optimal when considering genuine microscopic quantum systems. In addition, interferometric experiments that are suitable for genuine (microscopic) quantum systems involve coherent superpositions of two states, which could be, in the case of many-body macroscopic systems, experimentally infeasible with current technology. To confirm our analysis for the fidelity LE,  we also present quantitative results for the fidelity-induced first time derivative of the rate function in the case of the 1D Su-Schrieffer-Heeger (SSH) topological insulator and the 2D Massive Dirac (MD) model of a Chern insulator~\cite{qi:hug:zha:08}.\\  

\section{Dynamical (quantum) phase transitions and associated susceptibilities}
\label{Sec: DQPTs and Susceptibilities}
The authors in~\cite{hey:pol:keh:13} introduce the concept of DQPTs and illustrate their properties on the case of the transverse-field Ising model. They observe a similarity between the partition function of a quantum system in equilibrium, $Z(\beta)=\tr(e^{-\beta H})$, and the overlap amplitude of some time-evolved initial quantum state $\ket{\psi_i}$ with itself, $G(t)=\langle\psi_i|e^{-iHt}|\psi_i\rangle$. During a temperature-driven PT, the abrupt change of the properties of the system is indicated by the non-analyticity of the free energy density $f(\beta)=-\lim_{N\to\infty}\frac{1}{N}\ln Z(\beta)$ at the critical temperature ($N$ being the number of degrees of freedom). It is then possible to establish an analogy with the case of the real time evolution of a quantum system out of equilibrium, by considering the rate function 
\begin{equation}
\label{eq:rate}
g(t)=-\frac{1}{N}\log |G(t)|^2,
\end{equation} 
where $|G(t)|^2$ is a mixed-state LE, as we detail below. The rate function $g(t)$ may exhibit non-analyticities at some critical times $t_c$, after a quantum quench. This phenomenon is termed DPT. 

We study DPTs for mixed states using the fidelity and the interferometric LEs. We first investigate the relation between the two approaches for DQPTs at zero temperature. More concretely, we perform analytical derivations of the corresponding susceptibilities in the general case of a family of static Hamiltonians, parametrised by some smooth manifold $M$, $\{H(\lambda):\lambda\in M\}$. \\

\subsection{DQPTs for pure states}
\label{Subsec: DQPT for pure states}

At zero temperature, the LE $G(t)$ from~\eqref{eq:rate} between the ground state for $\lambda=\lambda_i\in M$ and the evolved state with respect to the Hamiltonian for $\lambda=\lambda_f\in M$ is given by the fidelity between the two states
\begin{eqnarray}
\mathcal{F}(t;\lambda_{f},\lambda_i)\equiv |\bra{\psi(\lambda_i)}e^{-it H(\lambda_f)}\ket{\psi(\lambda_i)}|.
\label{Eq: zero temperature fidelity}
\end{eqnarray}
For $\lambda_i=\lambda_f$, the fidelity is trivial, since the system remains in the same state. Fixing $\lambda_i\equiv \lambda$ and $\lambda_{f} = \lambda + \delta\lambda$, with $\delta\lambda<<1$, in the $t\rightarrow\infty$ limit Eq.~\eqref{Eq: zero temperature fidelity} is nothing but the familiar $S$-matrix with an unperturbed Hamiltonian $H(\lambda)$ and an interaction Hamiltonian $V(\lambda)$, which is approximated by
\begin{eqnarray}
V(\lambda)\equiv H(\lambda_f)-H(\lambda)\approx\frac{\partial H}{\partial \lambda^{a}}(\lambda)\delta \lambda^a.
\label{Eq:2}
\end{eqnarray}
After applying standard perturbation theory techniques (for details, see Appendix), we obtain
\begin{eqnarray}
\mathcal{F}(t;\lambda_f,\lambda)\approx 1 -\chi_{ab}(t;\lambda)\delta \lambda^a \delta \lambda^b,
\end{eqnarray}
where the dynamical susceptibility $\chi_{ab}(t;\lambda)$ is given by
\begin{small}
\begin{align}
\chi_{ab}(t;\lambda)=
\int_{0}^{t}\int_{0}^{t} dt_2dt_1 \left(\frac{1}{2}\langle \{V_{a}(t_2),V_{b}(t_1)\} \rangle-\langle V_{a}(t_2)\rangle \langle V_{b}(t_1) \rangle\right),
\label{Eq: zero temperature susceptibility}
\end{align}
\end{small}with $V_{a}(t,\lambda)=e^{it H(\lambda)}\partial H/\partial \lambda^{a}(\lambda)e^{-it H(\lambda)}$ and $\langle \ast\rangle=\bra{\psi(\lambda)}\ast\ket{\psi(\lambda)}$. The family of symmetric tensors $\{ds^2(t)=\chi_{ab}(t,\lambda)d\lambda^a d\lambda^b\}_{t\in \mathbb{R}}$ defines a family of metrics in the manifold $M$, which can be seen as pullback metrics of the Bures metric (Fubini-Study metric) in the manifold of pure states~\cite{chr:jam:12}. Specifically, at time $t$, the pullback is given by the map $\Phi_t: \lambda_f \mapsto e^{-it H(\lambda_f)}\ket{\psi(\lambda)}\bra{\psi(\lambda)}e^{i t H(\lambda_f)}$, evaluated at $\lambda_f=\lambda$.
 
\subsection{Generalizations at finite temperatures}
\label{Subsec: DQPT for pure states}

The generalization of DQPTs to mixed states is not unique. There are several ways to construct a LE for a general density matrix. In what follows, we derive two finite temperature generalizations, such that they have the same zero temperature limit. \\
\subsubsection{Fidelity Loschmidt Echo at $T>0$}

First, we introduce the {\em fidelity LE} between the state $\rho(\beta;\lambda_i)=e^{-\beta H(\lambda_i)}/\tr\{e^{-\beta H(\lambda_i)}\}$ and the one evolved by the unitary operator $e^{-it H(\lambda_f)}$ as
\begin{eqnarray}
\!\!\!\!\!\!\mathcal{F}(t,\beta;\lambda_i,\lambda_f)\!=\!F(\rho(\beta;\lambda_i),e^{-it H(\lambda_f)}\rho(\beta;\lambda_i)e^{itH(\lambda_f)}),
\label{eq:fid}
\end{eqnarray}
where $F(\rho,\sigma)=\tr\{\sqrt{\sqrt{\rho}\sigma\sqrt{\rho}}\}$ is the quantum fidelity between arbitrary mixed states $\rho$ and $\sigma$. For $\lambda_f$ close to $\lambda_i=\lambda$, we can write
\begin{eqnarray}
\mathcal{F}(t,\beta ;\lambda_f,\lambda)\approx 1 -\chi_{ab}(t,\beta ;\lambda)\delta \lambda^a \delta \lambda^b,
\end{eqnarray}
with $\chi_{ab}(t,\beta,\lambda)$ being the \emph{Dynamical Fidelity Susceptibility} (DFS). Notice that $\lim_{\beta\to\infty}\chi_{ab}(t,\beta;\lambda)=\chi_{ab}(t;\lambda)$, where $\chi_{ab}(t;\lambda)$ is given by Eq.~\eqref{Eq: zero temperature susceptibility}. At time $t$ and inverse temperature $\beta$, we have a map $\Phi_{(t,\beta)}: \lambda_f\mapsto e^{-i tH (\lambda_f)} \rho(\beta;\lambda) e^{it H(\lambda_f)}$. The $2$-parameter family of metrics defined by $ds^2(\beta,t)=\chi_{ab}(t,\beta;\lambda)d\lambda^{a}d\lambda^b$ is the pullback by $\Phi_{(t,\beta)}$ of the Bures metric on the manifold of full-rank density operators, evaluated at $\lambda_f=\lambda$ (see Appendix).

The fidelity LE is closely related to the Uhlmann connection: $F(\rho_1,\rho_2)$ equals the overlap between purifications $W_1$ and $W_2$, $\langle W_1,W_2\rangle=\tr \left\{W_1^\dagger W_2\right\}$, satisfying discrete parallel transport condition (see, for instance~\cite{uhl:11}). \\

\subsubsection{Interferometric Loschmidt Echo at $T>0$}

Here, we consider an alternative definition of the LE for mixed states [$G(t)$ from Eq.~\eqref{eq:rate}]. In particular, we define the {\em interferometric LE} as
\begin{eqnarray}
\mathcal{L}(t,\beta ;\lambda_{f},\lambda_i)=\left|\frac{\tr\left\{e^{-\beta H(\lambda_i)} e^{i tH(\lambda_i)}e^{-i t H(\lambda_f)}\right\}}{\tr \left\{e^{-\beta H(\lambda_i)}\right\}}\right|.
\end{eqnarray}
The $e^{it H(\lambda_i)}$ factor does not appear at zero temperature, since it just gives a phase which is canceled by taking the absolute value.  This differs from previous treatments in the literature~\cite{bud:hey:16} (see Section 5.5.4 of~\cite{chr:jam:12}, where the variation of the interferometric phase, $\tr\{\rho_0 e^{-it H}\}$, exposes this structure). However, it is convenient to introduce it in order to have the usual form of the perturbation expansion, as will become clear later.

For $\lambda_f$ close to $\lambda_i=\lambda$, we get
\begin{eqnarray}
\mathcal{L}(t,\beta ;\lambda_f,\lambda)\approx \left|\frac{\tr\left\{e^{-\beta H(\lambda)} T e^{-i \int_{0}^{t} dt' V_{a}(t,\lambda) \delta \lambda^a}\right\}}{\tr \left\{e^{-\beta H(\lambda)}\right\}}\right|,
\end{eqnarray}
so that the perturbation expansion goes as in Eq.~\eqref{Eq: zero temperature susceptibility}, yielding
\begin{eqnarray}
\mathcal{L}(t,\beta ;\lambda_f,\lambda)\approx 1 -\tilde\chi_{ab}(t,\beta ;\lambda)\delta \lambda^a \delta \lambda^b,
\end{eqnarray}
with the dynamical susceptibility given by
\begin{small}
\begin{align}
\tilde{\chi}_{ab}(t,\beta;\lambda)=
 \int_{0}^{t}\int_{0}^{t} dt_2dt_1  \left(\frac{1}{2}\langle \{V_{a}(t_2),V_{b}(t_1)\} \rangle-\langle V_{a}(t_2)\rangle \langle V_{b}(t_1) \rangle\right),
\label{Eq: finite temperature interferometric susceptibility}
\end{align}
\end{small}where $\langle \ast\rangle=\tr\{e^{-\beta H(\lambda)}\ast\}/\tr\{e^{-\beta H(\lambda)}\}$. Notice that Eqs~\eqref{Eq: finite temperature interferometric susceptibility} and~\eqref{Eq: zero temperature susceptibility} are formally the same with the average over the ground state replaced by the thermal average. This justifies the extra $e^{it H(\lambda_i)}$ factor. Since this susceptibility comes from the interferometric LE, we call it \emph{Dynamical  Interferometric Susceptibility} (DIS). The quantity $ds^2(\beta,t)=\tilde{\chi}_{ab}(t,\beta;\lambda)d\lambda^ad\lambda^b$ defines a $2$-parameter family of metrics over the manifold $M$, except that they cannot be seen as pullbacks of metrics on the manifold of density operators with full rank. However, it can be interpreted as the pullback by a map from $M$ to the unitary group associated with the Hilbert space of a particular Riemannian metric. For a detailed analysis, see Appendix. Additionally, we point out that this version of LE is related to the interferometric geometric phase introduced by Sj\"{o}qvist \emph{ et. al}~\cite{sjo:pat:eke:ana:eri:oi:ved:00,ton:sjo:kwe:oh}. \\

\subsection{Two-band systems}

Many representative examples of topological insulators and superconductors can be described by effective two-band Hamiltonians. Therefore, we derive closed expressions of the previously introduced dynamical susceptibilities for topological systems within this class. 

The general form of such Hamiltonians is $\{H(\lambda)=\vec{x}(\lambda)\cdot \vec\sigma:\lambda \in M\}$, where $\vec{\sigma} $  is the Pauli vector. The interaction Hamiltonian $V(\lambda)$, introduced in Eq.~\eqref{Eq: zero temperature fidelity}, casts the form
\begin{align*}
V(\lambda)\approx \left(\frac{\partial \vec{x}}{\partial\lambda^a}\cdot \vec{\sigma}\right) \delta \lambda^a.
\end{align*} 
It is convenient to decompose $\partial\vec{x}/\partial \lambda^a$ into one component perpendicular to $\vec{x}$ and one parallel to it:
\begin{align*}
\frac{\partial \vec{x}}{\partial\lambda^a}=\left(\frac{\partial \vec{x}}{\partial\lambda^a}\right)^{\perp}+ \left(\frac{\partial \vec{x}}{\partial\lambda^a}\right)^{\parallel}=\vec{t}_{a}+\vec{n}_a.
\end{align*}
The first term is tangent, in $\mathbb{R}^3$, at $\vec{x}(\lambda)$, to a sphere of constant radius $r=|\vec{x}(\lambda)|$. Hence, this kind of perturbations does not change the spectrum of $H$, only its eigenbasis. The second term is a variation of the length of $\vec{x}$ and hence, it changes the spectrum of $H$, while keeping the eigenbasis fixed. The DFS and the DIS are given by (for the details of the derivation, see Appendix)
\begin{small}
\begin{eqnarray}
\label{Eq: DFS}
\chi_{ab}&=&\tanh^2(\beta |\vec{x}(\lambda)|)\frac{\sin^2(|\vec{x}(\lambda)|t)}{|\vec{x}(\lambda)|^2}\vec{t}_a\!\!\cdot\!\vec{t}_b \\
\label{Eq: DIS}
\tilde{\chi}_{ab}\!&=&\!\frac{\sin^2(|\vec{x}(\lambda)|t)}{|\vec{x}(\lambda)|^2}\vec{t}_a\!\!\cdot\!\vec{t}_b\!+\!t^2(1\!\!-\!\!\tanh^2(\beta |\vec{x}(\lambda)|))\vec{n}_a\!\!\!\cdot\!\vec{n}_b.
\end{eqnarray}
\end{small}

While the DIS~\eqref{Eq: DIS} depends on the variation of both the spectrum and the eigenbasis of the Hamiltonian, the DFS~\eqref{Eq: DFS} depends only on the variations which preserve the spectrum, i.e., changes in the eigenbasis. This is very remarkable, in general the fidelity between two quantum states, being their distinguishability measure, does depend on both the variations of the spectrum and the eigenbasis. In our particular case of a quenched system, the eigenvalues are preserved (see Eq.~\eqref{eq:fid}), as the system is subject to a unitary evolution. The tangential components of both susceptibilities are modulated by the function $\sin ^2 (Et)/ E^2$, where $E$ is the gap. This captures the \emph{Fisher zeros}, i.e., the zeros of the (dynamical) partition function which here is given by the fidelity $\mathcal F$ from~\eqref{eq:fid} (see~\cite{yan:lee:52,lee:yan:52, fis:65}). Observe that whenever $t=(2n+1)\pi/2E$, $n\in\mathbb{Z}$, this factor is maximal and hence, both LEs decrease abruptly.
The difference between the two susceptibilities is given by
\begin{align*}
\tilde{\chi}_{ab}-\chi_{ab}
=(1-\tanh^2(\beta|\vec{x}(\lambda)|))\left(\frac{\sin^2(|\vec{x}(\lambda)|t)}{|\vec{x}(\lambda)|^2}\vec{t}_a\cdot \vec{t}_{b} +t^2\vec{n}_a\cdot \vec{n}_b\right).
\end{align*}
The quantity $(1-\tanh^2(\beta E))$ is nothing but the static susceptibility, see~\cite{pat:96}. Therefore, the difference between DIS and DFS is modulated by the static susceptibility at finite temperature.

To illustrate the relationship between the two susceptibilities, in FIG. 1 we plotted the modulating function for the tangential components of both. We observe that at zero temperature they coincide. As the temperature increases, in the case of the fidelity LE, the gap-vanishing points become less important. On the contrary, for the interferometric LE, the associated tangential part of the susceptibility does not depend on temperature, thus the gap-vanishing points remain prominent. The DFS from Eq.~\eqref{Eq: DFS} thus predicts gradual smearing of critical behaviour, consistent with previous findings that showed the absence of phase transitions at finite temperatures in the static case~\cite{mer:vla:pau:vie:17,mer:vla:pau:vie:17:qw}. The DIS from Eq.\eqref{Eq: DIS} has a tangential term that is not coupled to the temperature, persisting at higher temperatures and giving rise to abrupt changes in the finite-temperature system's behaviour. This is also consistent with previous studies in the literature, where DPTs were found even at finite temperatures~\cite{hey:bud:17, bah:ban:dut:17}. Additionally, the interferometric LE depends on the normal components of the variation of $\vec{x}$. In other words, the finite-temperature phase transitions inferred by the behavior of the interferometric LE occur due to the change of the parameters of the Hamiltonian and not due to temperature. \\

\begin{figure}[h]
    \includegraphics[scale=0.4]{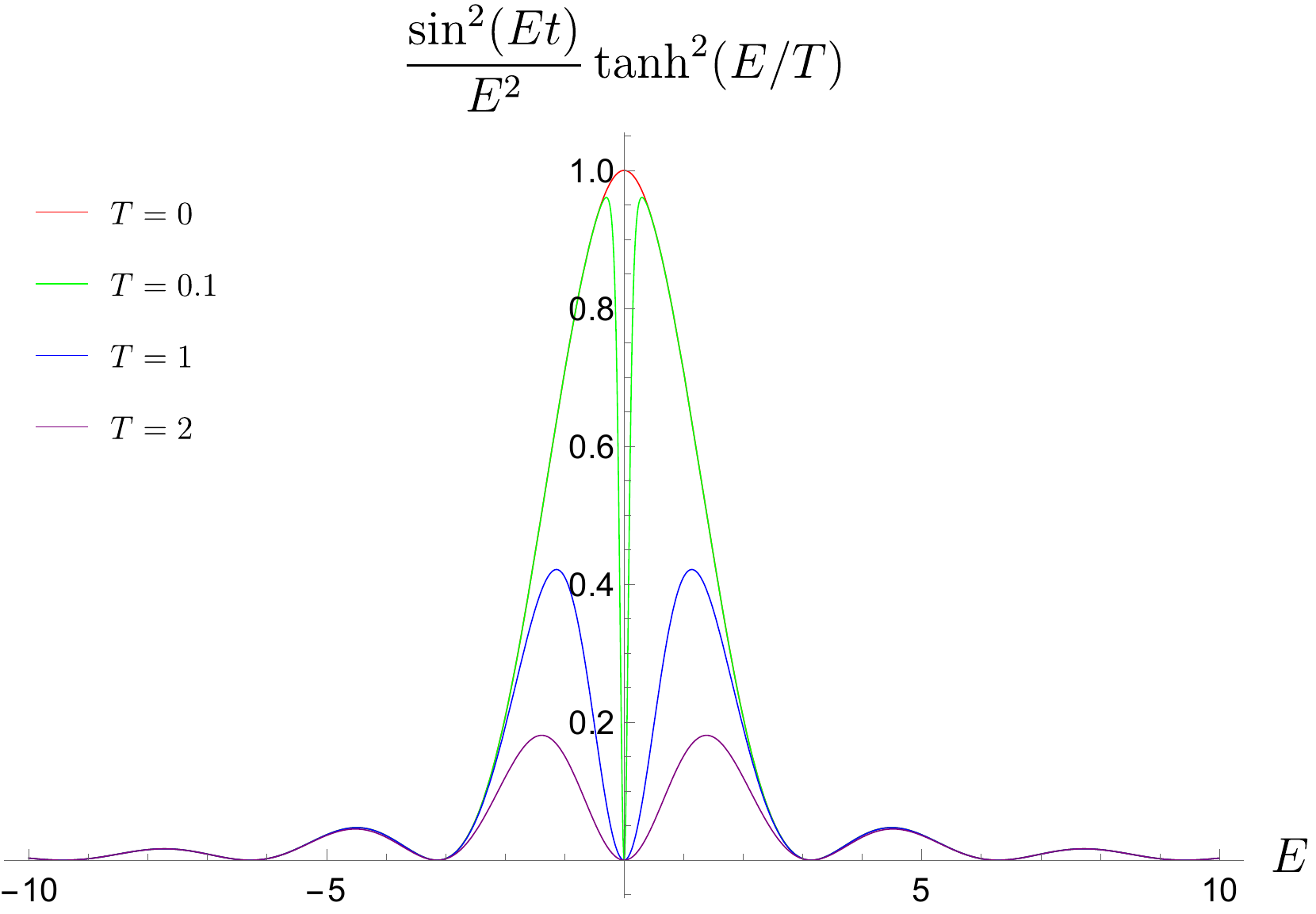}
    \caption{The susceptibility modulating function for the tangential components at $t=1$.}
\end{figure}

\subsection{Comparing the two approaches}

The above analysis of the two dynamical susceptibilities (metrics) reflects the essential difference between the two distinguishability measures, one based on the fidelity, the other on  interferometric experiments. From the quantum information theoretical point of view, the two quantities can be interpreted as distances between \emph{states}, or between \emph{processes}, respectively. The Hamiltonian evaluated at a certain point of parameter space $M$ defines the macroscopic phase. Associated to it we have thermal states and unitary processes. The fidelity LE is obtained from the Bures distance between a thermal state $\rho_1$ in phase $1$ and the one obtained by unitarily evolving this state, $\rho_2 = U_2\rho_1 U_2^{\dagger}$, with $U_2$ associated to phase $2$.  Given a thermal state $\rho_1$ prepared in phase $1$, the interferometric LE is obtained from the distance between two unitary processes $U_1$ and $U_2$ (defined modulo a phase factor), associated to phases $1$ and $2$.  

The quantum fidelity between two states is in fact the classical fidelity between the probability distributions obtained by performing an optimal measurement on them. Measuring an observable $M$ on the two states $\rho_1$ and $\rho_2$, one obtains the probability distributions $\{p_1(i)\}$ and $\{p_2(i)\}$, respectively. The quantum fidelity $F_Q$ between the two states $\rho_1$ and $\rho_2$ is bounded by the classical fidelity $F_C$ between the probability distributions $\{p_1(i)\}$ and $\{p_2(i)\}$, $F_Q (\rho_1,\rho_2) = \mbox{Tr}\sqrt{\sqrt{\rho_1}\rho_2\sqrt{\rho_1}} \leq \sum_i \sqrt{p_1(i)p_2(i) = F_c (p_1(i),p_2(i))}$, such that the equality is obtained by measuring an {\em optimal observable}, given by $M_{\text{op}} = \rho_1^{-1/2}\sqrt{\sqrt{\rho_1}\rho_2\sqrt{\rho_1}} \rho_1^{-1/2}$ (note that optimal observable is not unique). For that reason, one can argue that the fidelity is capturing all order parameters (i.e., measurements) through its optimal observables $M_{\text{op}}$. Fidelity-induced distances, the {\em Bures distance} $D_B(\rho_1,\rho_2) = \sqrt{2(1-F_Q(\rho_1,\rho_2))}$, the {\em sine distance} $D_S(\rho_1,\rho_2) = \sqrt{1-F_Q^2(\rho_1,\rho_2)}$ and the {\em F-distance} $D_F(\rho_1,\rho_2) = 1-F_Q(\rho_1,\rho_2)$ satisfy the following set of inequalities
\begin{equation*}
	 D_F(\rho_1,\rho_2) \leq D_T(\rho_1,\rho_2) \leq D_S(\rho_1,\rho_2) \leq D_B(\rho_1,\rho_2),
\end{equation*}
where the {\em trace distance} is given by $D_T(\rho_1,\rho_2) = \frac{1}{2} \mbox{Tr}|\rho_1 - \rho_2|$. In other words, the fidelity-induced distances and the trace distance establish the same order on the space of quantum states. This is important, as the trace distance is giving the optimal value for the success probability in ambiguously discriminating in a {\em single-shot measurement} between two {\em a priori} equally probable states $\rho_1$ and $\rho_2$, given by the so-called Helstrom bound $P_H (\rho_1,\rho_2) = (1 + D_T (\rho_1,\rho_2))/2$~\cite{hel:76}.

On the other hand, the interferometric phase is based on some interferometric experiment to distinguish two states, $\rho_1 = \sum_i r_i \ket{i}\bra{i}$ and $\rho_2 = U_2 \rho_1 U_2^{\dagger}$: it measures how the intensities at the outputs of the interferometer are affected by applying $U_2$ to only one of its arms~\cite{sjo:pat:eke:ana:eri:oi:ved:00}. Therefore, to set up such an experiment, one does not need to know the state $\rho_1$ that enters the interferometer, as only the knowledge of $U_2$ is required. Note that this does not mean that the output intensities do not depend on the interferometric LE: indeed, the inner product $\langle U_1,U_2\rangle_{\rho_1}$ is defined with respect to the state $\rho_1$. This is a different type of experiment, not based on the observation of any physical property of a system. It is analogous to comparing two masses with weighing scales, which would show the same difference of $\Delta m = m_1 - m_2$, regardless of how large the two masses $m_1$ and $m_2$ are. For that reason, interferometric distinguishability is more sensitive than the fidelity (fidelity depends on more information, not only how much the two states are different, but in what aspects this difference is observable). Indeed,  the interferometric LE between $\rho_1$ and $\rho_2$ can be written as the overlap $L(\rho_1, \rho_2) = |\langle \rho_1|\rho_2\rangle|$ between the purifications $|\rho_1\rangle = \sum_i  \sqrt{r_i} |i\rangle|i\rangle$ and $|\rho_2\rangle=(U\otimes I)|\rho_1\rangle$. On the other hand, the fidelity satisfies $F(\rho_1, \rho_2) = \max_{|\psi\rangle,|\varphi\rangle} |\langle\psi|\varphi\rangle|$, where $|\psi\rangle$ and $|\varphi\rangle$ are purifications of $\rho_1$ and $\rho_2$, respectively, i.e., $L(\rho_1, \rho_2) \leq F(\rho_1, \rho_2)$. Moreover, what one does observe in interferometric experiments are the mentioned output intensities, i.e., one needs a number of identical systems prepared in the same state to obtain results in interferometric measurements. This additionally explains why interferometric LE is more sensible than the fidelity one, as the latter is based on the observations performed on single systems. The fact that interferometric LE is more sensitive than the fidelity LE is consistent with the result that the former is able to capture the changes of some of the system's features at finite temperatures (thus they predict DPTs), while the latter cannot.

In terms of experimental feasibility, the fidelity is more suitable for the study of many-body macroscopic systems and phenomena, while the interferometric measurements provide a more detailed information on genuinely quantum (microscopic) systems. Finally, interferometric experiments involve coherent superpositions of two states. Therefore, when applied to many-body systems, one would need to create genuine Schr\"{o}dinger cat-like states, which goes beyond the current, and any foreseeable, technology (and could possibly be forbidden by more fundamental laws of physics; see for example objective collapse theories~\cite{bas:loc:sat:sin:ulb:13}). \\  

\section{DPTs of topological insulators at finite temperatures}
\label{sec:num.results}
Our general study of two-band Hamiltonians showed that the fidelity-induced LE predicts a gradual smearing of DPTs with temperature. In order to test this result, we study the fidelity LE on concrete examples of two topological insulators (as noted in the main text, the analogous study for the interferometric LE on concrete examples has already been performed, and is consistent with our findings~\cite{hey:bud:17,bah:ban:dut:17}). In particular, we present quantitative results obtained for the first derivative of the rate function, $dg/dt$, where $g(t)=-\frac{1}{N}\log \mathcal{F}$, and 
\begin{eqnarray}
\!\!\!\!\!\!\mathcal{F}(t,\beta;\lambda_i,\lambda_f)\!=\!F(\rho(\beta;\lambda_i),e^{-it H(\lambda_f)}\rho(\beta;\lambda_i)e^{itH(\lambda_f)}). \nonumber
\end{eqnarray}

The fidelity $F$ is obtained by taking the product of the single-mode fidelities, each of  which has the form
\begin{widetext}
\begin{align*}
F(\rho(\beta;\lambda_i),e^{-it H(\lambda_f)}\rho(\beta;\lambda_i)e^{itH(\lambda_f)}) = \sqrt{\frac{1+\cosh^2(\beta E_i)+\sinh^2(\beta E_i)\left[\cos(2E_f t)+(1-\cos(2E_f t))(\vec{n}_i\cdot\vec{n}_f)^2)\right]}{2\cosh^2(\beta E_i)}},	
\end{align*}\end{widetext}
with $H_a=E_a\vec{n}_a\cdot \vec{\sigma}$ and $a=i,f$. This expression can be obtained by using Eq.~\eqref{eq:su(2)toso(3)} and the result found in the Supplemental Material of~\cite{mer:vla:pau:vie:17}. The quantity $dg/dt$ is the figure of merit in the study of the DQPTs, therefore we present the respective results that confirm the previous study: the generalisation of the LE with respect to the fidelity shows the absence of finite temperature dynamical PTs.  We consider two paradigmatic models of topological insulators, namely the SSH~\cite{su:sch:hee:79} and the MD~\cite{qi:hug:zha:08} models.

\subsection{SSH model (1D)}
The SSH model was introduced in~\cite{su:sch:hee:79} to describe polyacetilene, and it was later found to describe diatomic polymers~\cite{ric:mel:82}. In momentum space, the Hamiltonian for this model is of the form $H(k,m)=\vec{x}(k,m)\cdot\vec{\sigma},$ with $m$ being the parameter that drives the static PT. The vector $\vec{x}(k,m)$ is given by:
\begin{align*}
\vec{x}(k,m)=(m+\cos(k),\sin(k),0).
\end{align*}

By varying $m$ we find two distinct topological regimes. For $m<m_c=1$ the system is in a non-trivial phase with winding number $1$, while for $m>m_c=1$ the system is in a topologically trivial phase with winding number $0$.

\begin{figure}[h]
    \includegraphics[scale=0.3]{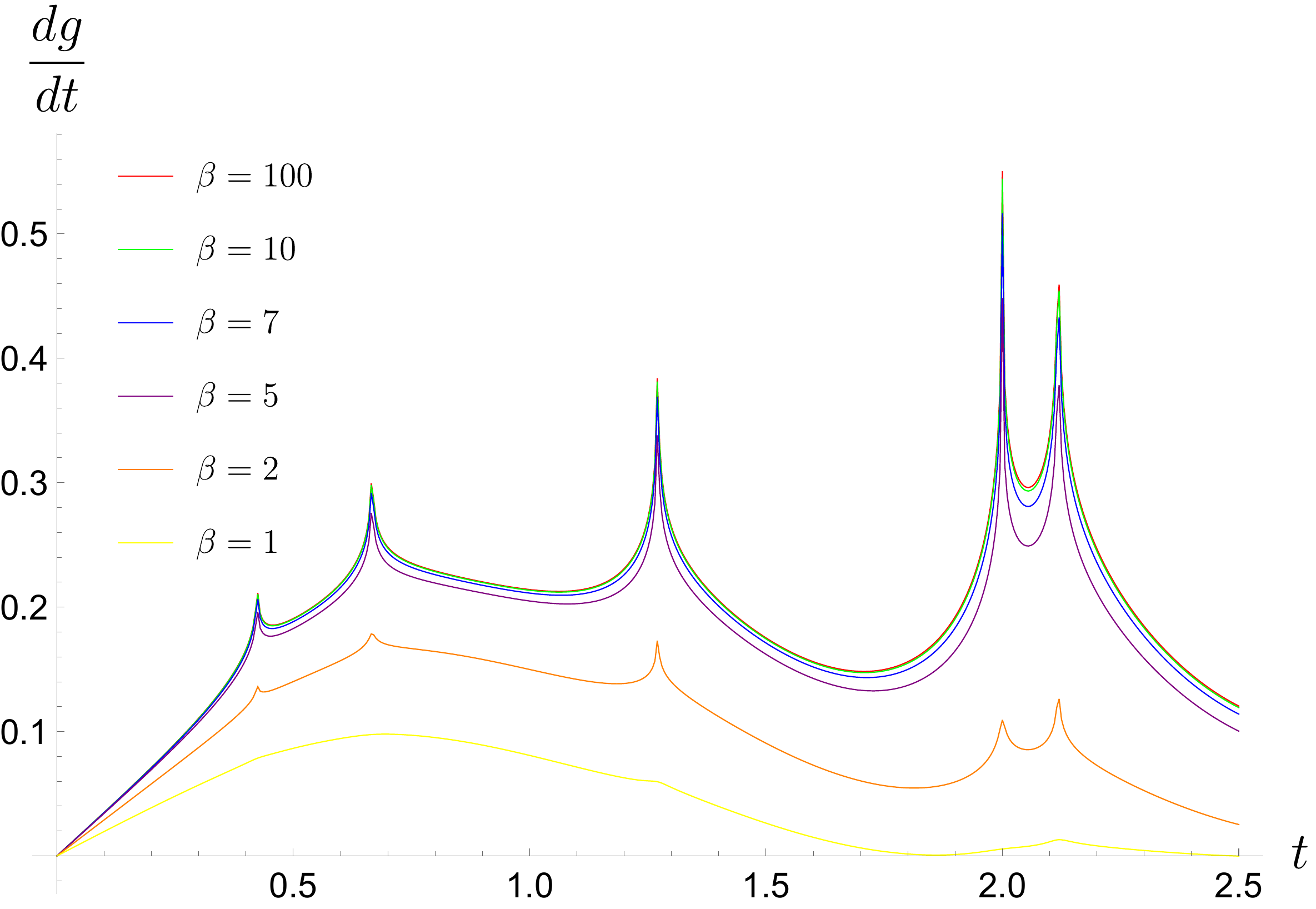}
    \caption{We plot the time derivative of the rate function, $dg/dt$, as a function of time for different values of the inverse temperature $\beta=1/T$. We consider a quantum quench from a trivial phase $(m=1.2)$ to a topological phase $(m=0.8)$.}
   \label{fig:SSH-trivial-to-topological}

\end{figure}

\begin{figure}[h]

\includegraphics[scale=0.3]{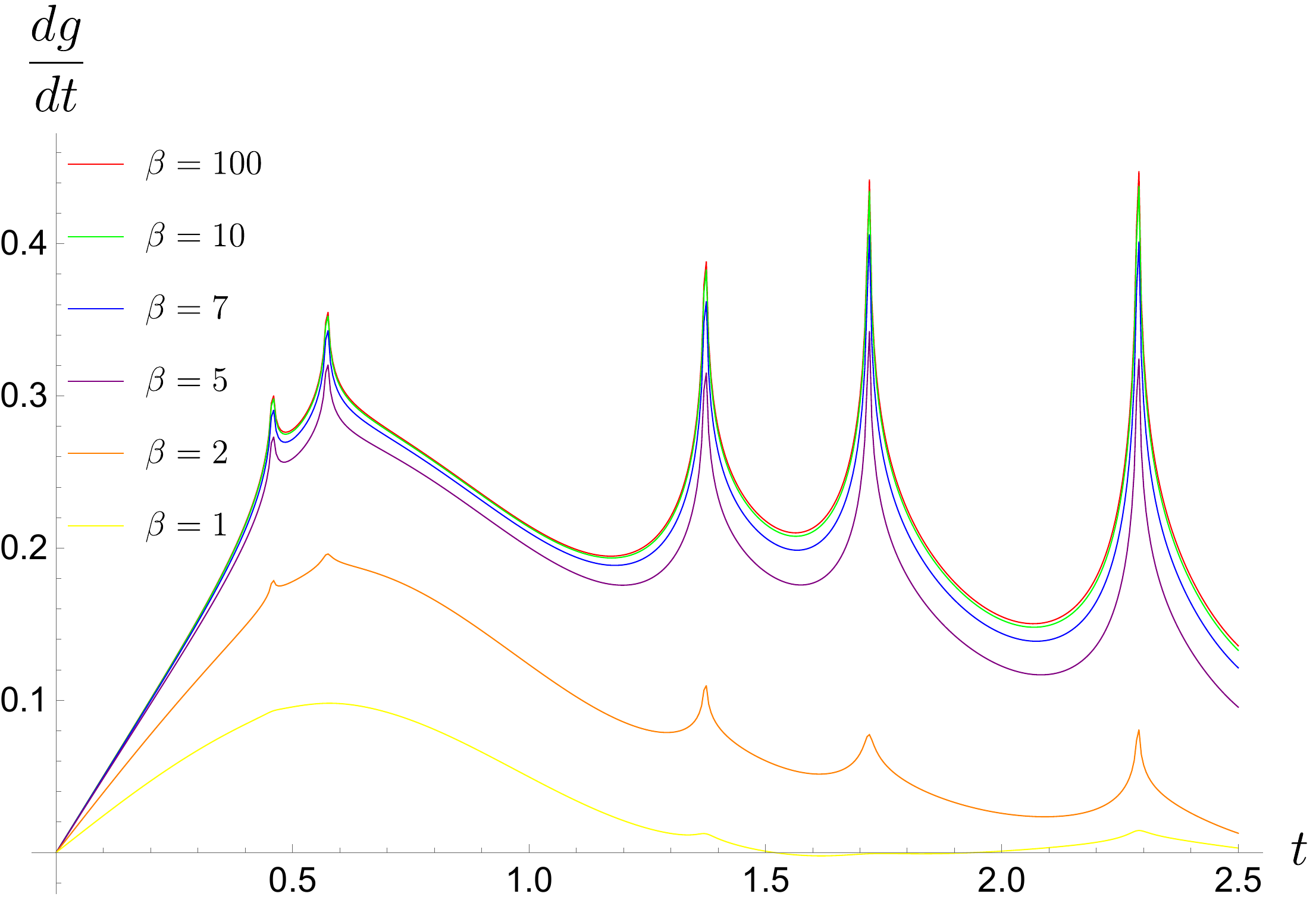}
    \caption{The time derivative of the rate function, $dg/dt$, as a function of time for different values of the inverse temperature. The quench is from a topological $(m=0.8)$ to a trivial phase $  (m=1.2)$.}

\label{fig:SSH-topological-to-trivial}
\end{figure}

We consider both cases in which we go from a trivial to a topological phase and vice versa (FIGs~\ref{fig:SSH-trivial-to-topological} and~\ref{fig:SSH-topological-to-trivial}, respectively). We notice that non-analyticities of the first derivative appear at zero temperature, and they are smeared out for higher temperatures.


\subsection{MD model (2D)}

The Massive Dirac model (MDM) captures the physics of a 2D Chern insulator~\cite{qi:hug:zha:08}, and shows different topologically distinct phases. In momentum space, the Hamiltonian for the MDM is of the form $H(\vec{k},m)=\vec{x}(\vec{k},m)\cdot\vec{\sigma},$ with $m$ being the parameter that drives the static PT. The vector $\vec{x}(\vec{k},m)$ is given by
\begin{align*}
\vec{x}(\vec{k},m)=(\sin(k_x),\sin(k_y),m-\cos(k_x)-\cos(k_y)).
\end{align*}
By varying $m$ we find four different topological regimes:

\begin{itemize}
\item For $-\infty<m<m_{c_1}=-2$ it is trivial (the Chern number is zero) -- Regime I
\item For $-2=m_{c_1}<m<m_{c_2}=0$ it is topological (the Chern number is $-1$) -- Regime II
\item For $0=m_{c_2}<m<m_{c_3}=2$ it is topological (the Chern number is $+1$) -- Regime III
\item For $2=m_{c_3}<m<\infty$ it is trivial (the Chern number is zero) -- Regime IV
\end{itemize}

\begin{figure}[h]

    \includegraphics[scale=0.3]{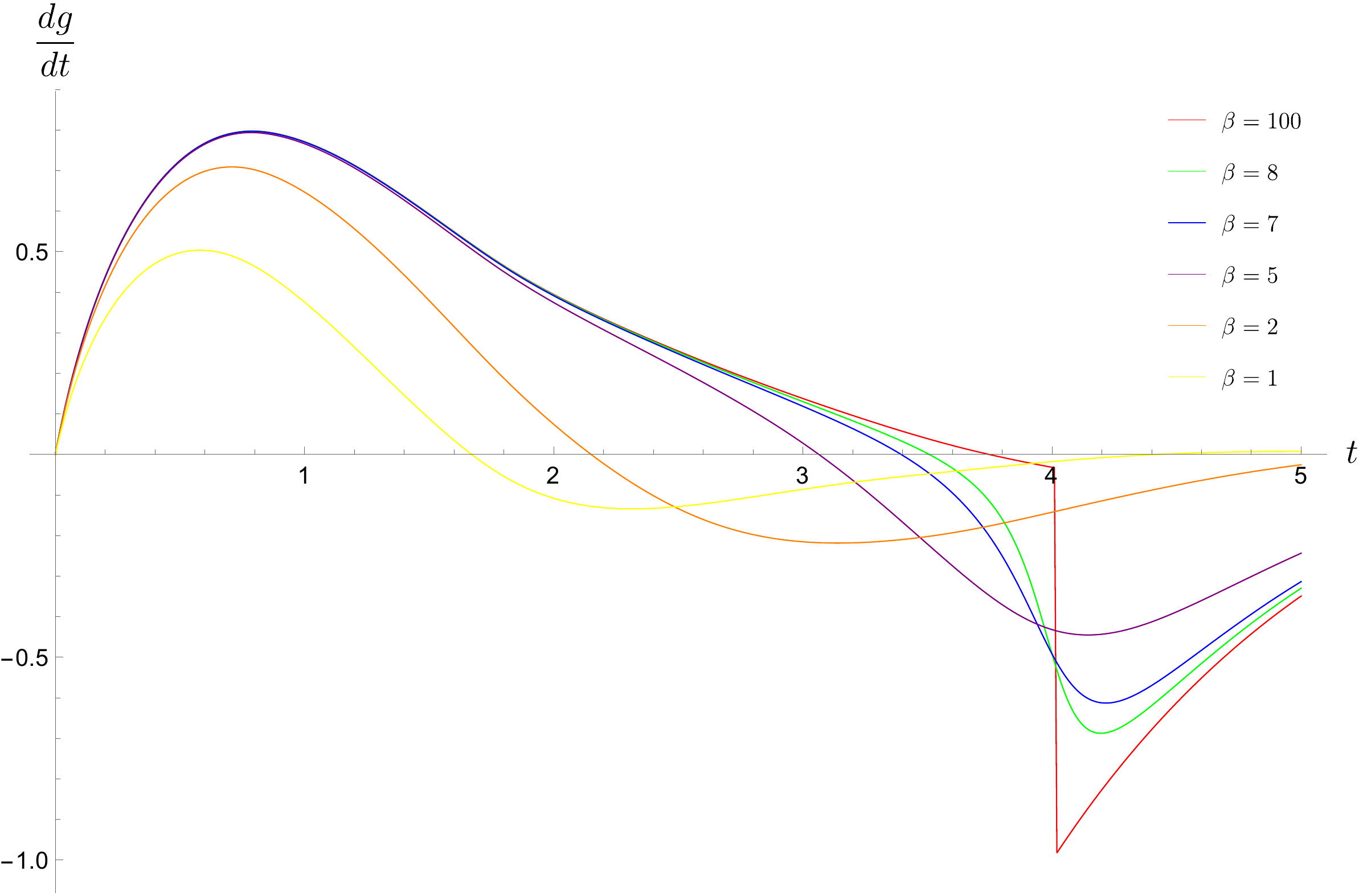}
    \caption{The time derivative of the rate function, $dg/dt$, as a function of time for different values of the inverse temperature. We quench the system from a trivial to a topological regime (Regimes from I to II and from IV to III).}
    \label{fig:MDM-trivial-L-topo--1}
    \end{figure}

\begin{figure}[h]    
    \includegraphics[scale=0.3]{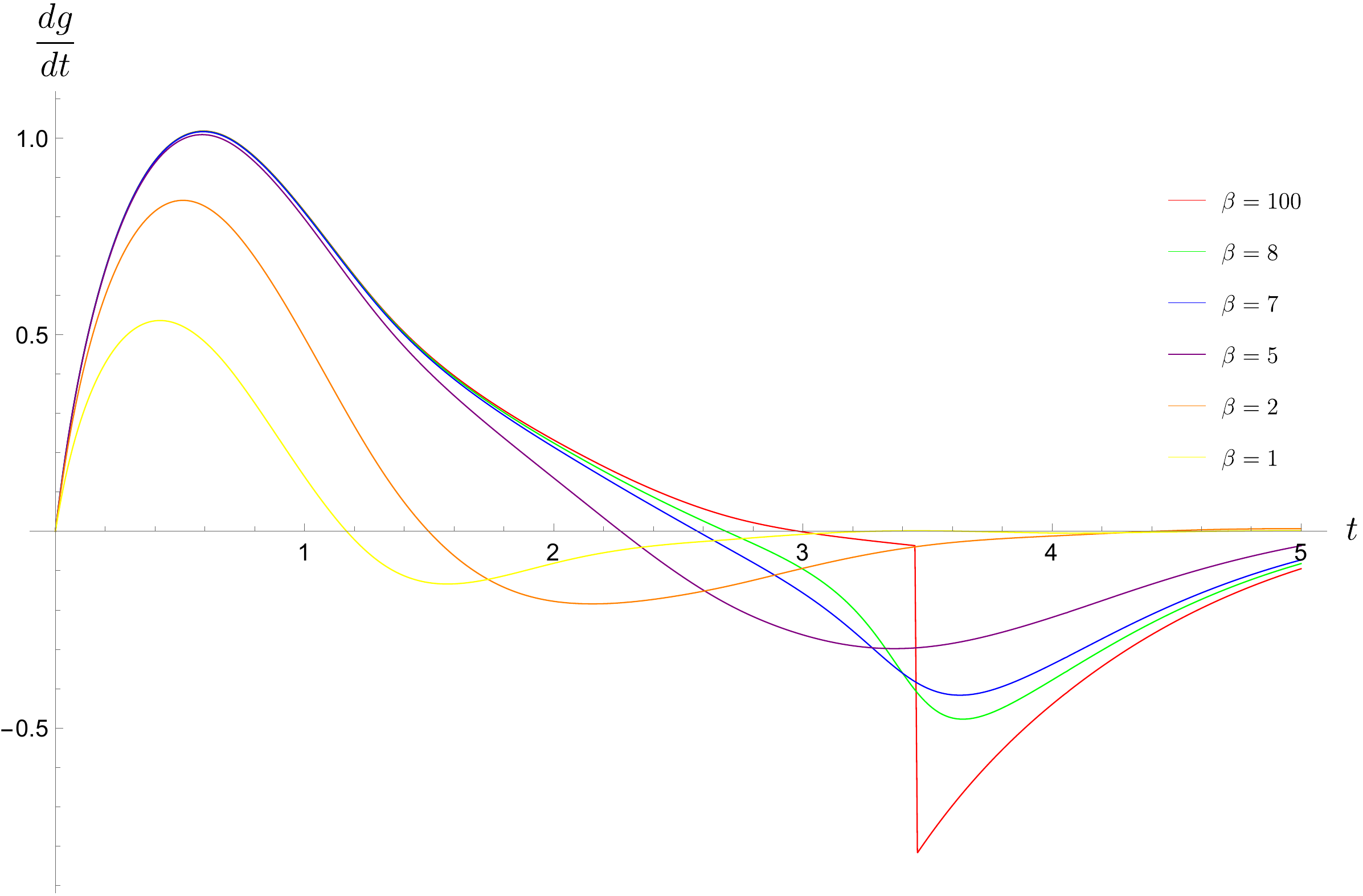}
    \caption{The time derivative of the rate function, $dg/dt$, as a function of time for different values of the inverse temperature. The quench is from a topological to a trivial regime (Regimes from II to I and from III to IV).}
     \label{fig:MDM-topo--1-trivial-L}
    \end{figure}

\begin{figure}[h]    
    \includegraphics[scale=0.3]{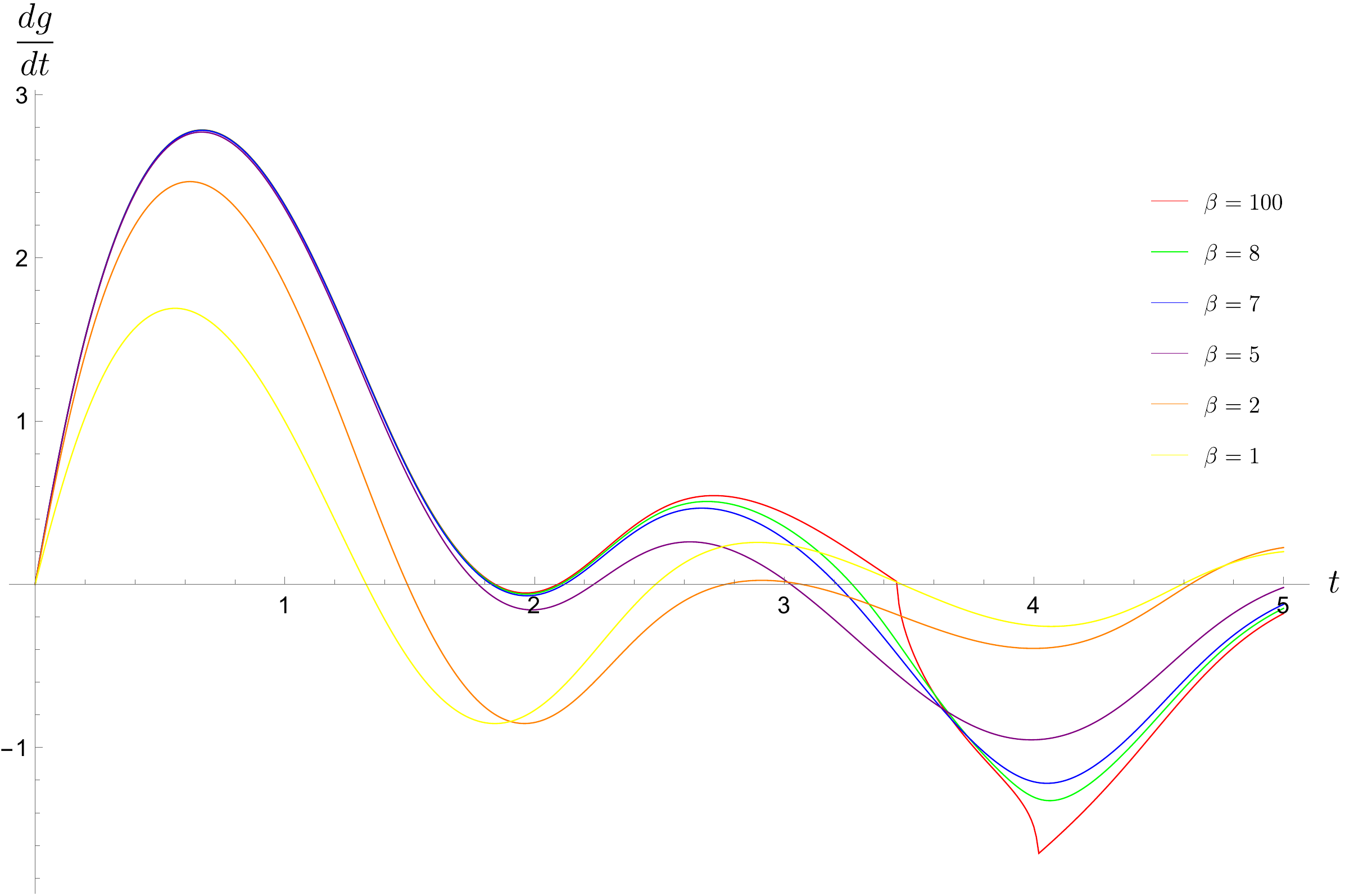}
    \caption{The time derivative of the rate function, $dg/dt$, as a function of time for different inverse temperatures. The quantum quench is from a topological to a topological regime (Regimes from II to III and vice versa).}
    \label{fig:MDM-topo-+1-topo--1}

\end{figure}

In FIGs~\ref{fig:MDM-trivial-L-topo--1},~\ref{fig:MDM-topo--1-trivial-L} and~\ref{fig:MDM-topo-+1-topo--1} we plot the first derivative of the rate function $g(t)$, as a function of time for different temperatures. We only consider quenches that traverse a single phase transition point. 

We observe that at zero temperature  there exist non-analyticities at the critical times -- the signatures of DQPTs. As we increase the temperature, these non-analyticities are gradually smeared out, resulting in smooth curves for higher finite temperatures. We note that the peak of the derivative $dg(t)/dt$ is drifted when increasing the temperature, in analogy to the usual drift of non-dynamical quantum phase transitions at finite temperature\cite{kem:que:smi:16}.

\begin{figure}[h!]
    \includegraphics[scale=0.3]{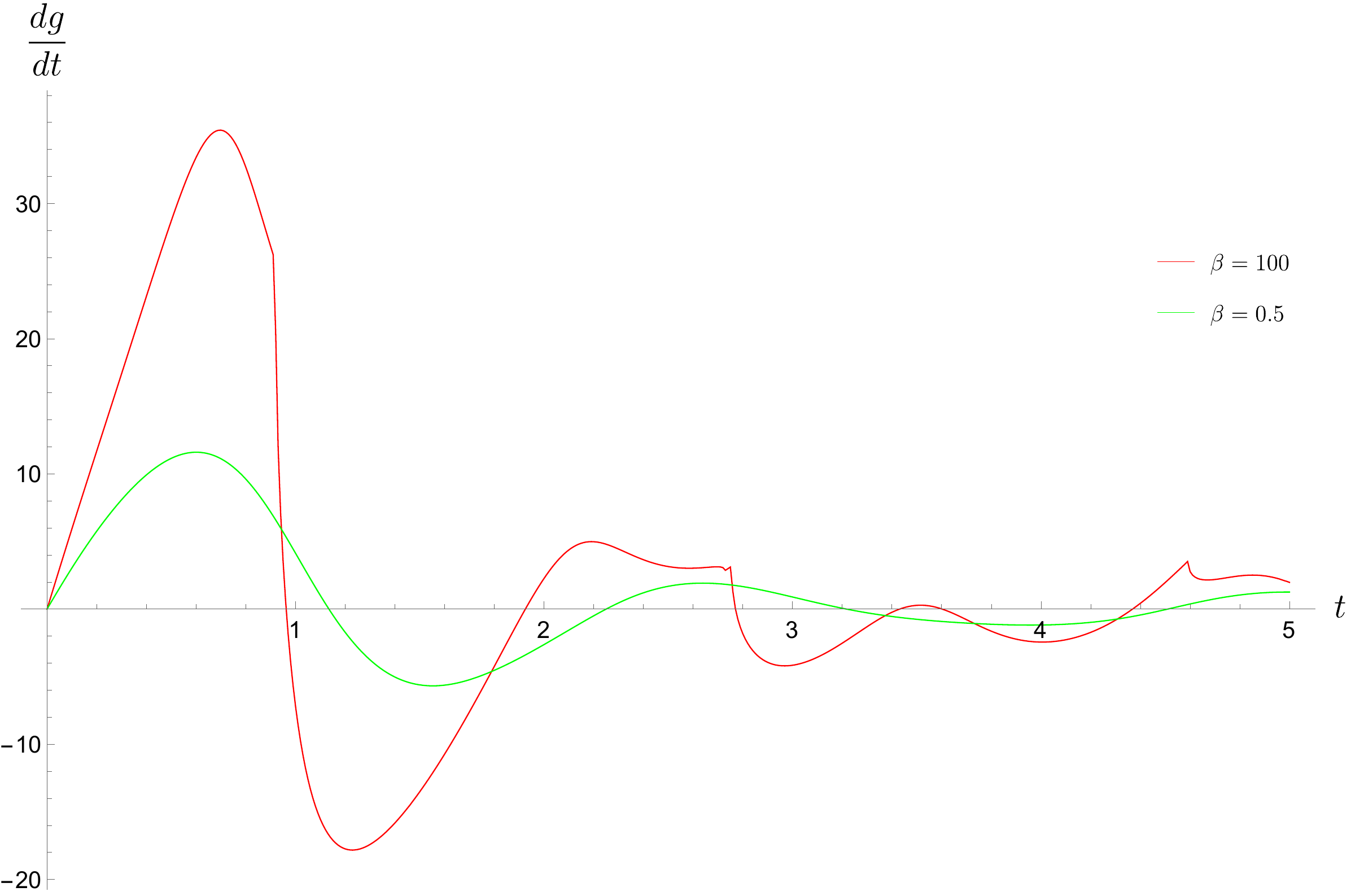}
    \caption{The time derivative of the rate function, $dg/dt$, as a function of time for different values of the inverse temperature, in the case that we quench the system from a trivial to a topological regime (Regimes from I to III and from IV to II).}
\label{fig:MDM-trivial-L-topo-+1}
\end{figure}

\begin{figure}[h]

    \includegraphics[scale=0.3]{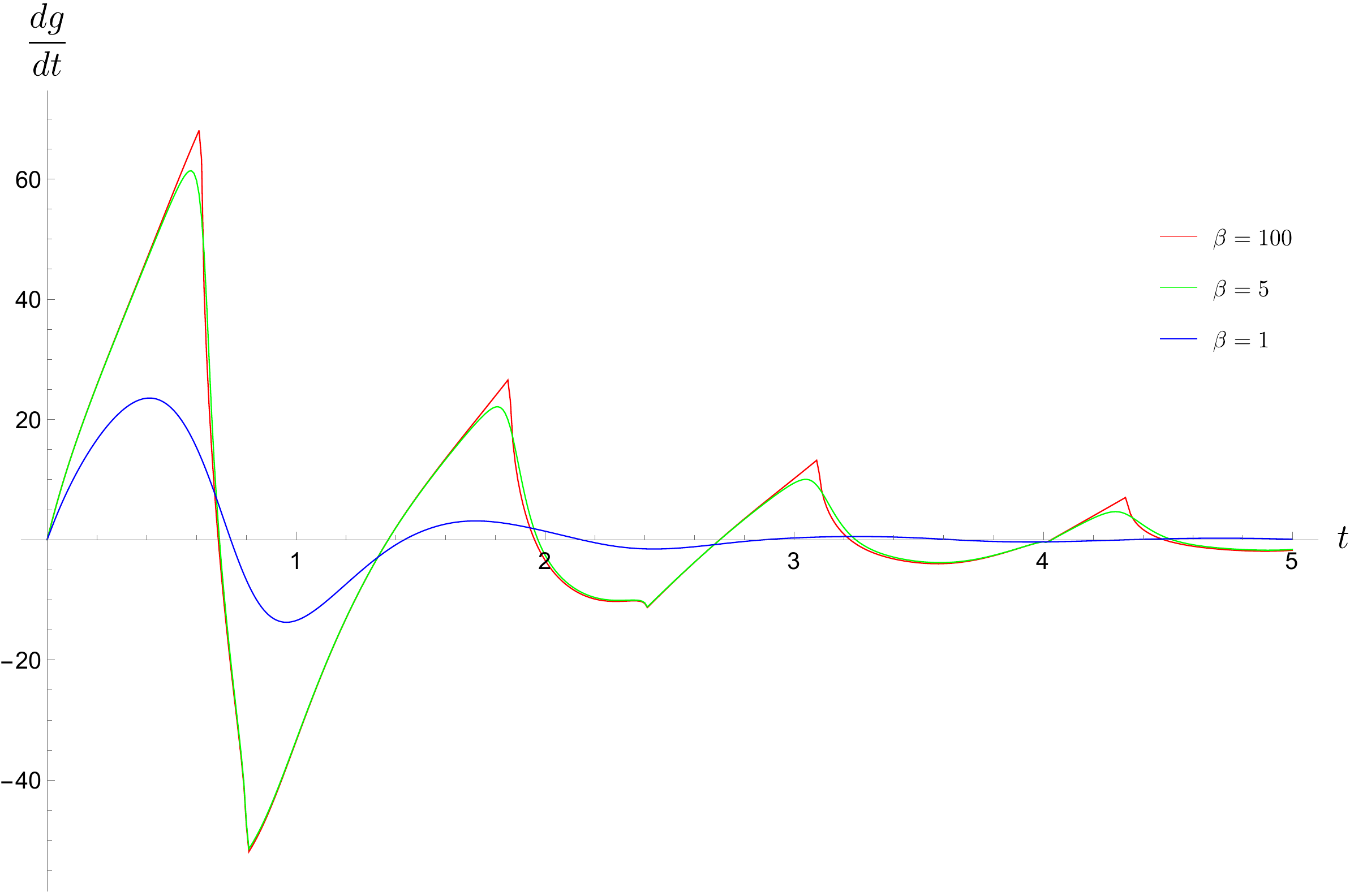}
    \caption{The time derivative of the rate function, $dg/dt$, as a function of time for different values of the inverse temperature. The system is quenched from a topological to a trivial regime (Regimes from III to I and from II to IV).}
\label{fig:MDM-topo--1-trivial-R}
\end{figure}

Next, we proceed by considering the cases in which we cross two phase transition points, as shown in FIGs~\ref{fig:MDM-trivial-L-topo-+1} and~\ref{fig:MDM-topo--1-trivial-R}. At zero temperature we obtain a non-analytic behaviour, which gradually disappears for higher temperatures.

Finally, we have also studied the case in which we move inside the same topological regime from left to right and vice versa. We obtained smooth curves without non -- analyticities, which we omit for the sake of briefness.

\section{Conclusions}

We analysed the fidelity and the interferometric generalisations of the LE for general mixed states, and applied them to the study of finite temperature DPTs in topological systems. 
We showed that the dynamical fidelity susceptibility is the pullback of the Bures metric in the \emph{space of density matrices}, i.e., states. On the other hand, the dynamical interferometric susceptibility is the pullback of a metric in the \emph{space of unitaries} (i.e., quantum channels). 

The difference between the two metrics reflects the fact that the fidelity is a measure of the state distinguishability between two {\em given} states $\rho$ and $\sigma$ in terms of observations, while the ``interferometric distinguishability'' quantifies how a quantum channel (a unitary $U$) changes an {\em arbitrary} state $\rho$ to $U \rho U^{\dagger}$. 
 
Therefore, while the ``interferometric distinguishability'' is in general more sensitive, and thus appropriate for the study of genuine (microscopic) systems, it is the fidelity that is the most suitable for the study of many-body system phases.
 
 Moreover, interferometric experiments involve coherent superpositions of two states, which for many-body systems would require creating and manipulating genuine Schr\"{o}dinger cat-like states. This seems to be experimentally beyond current technology.

We presented analytic expressions for the dynamical susceptibilities in the case of two-band Hamiltonians. At finite temperature, the fidelity LE indicates gradual disappearance of the zero-temperature DQPTs, while the interferometric LE predicts finite-temperature DPTs. We have performed finite temperature study on two representatives of topological insulators: the 1D SSH and the 2D MD models. In perfect agreement with the general result, the fidelity-induced first derivatives gradually smear down with temperature, not exhibiting any critical behaviour at finite temperatures. This is consistent with recent studies of 1D symmetry protected topological phases at finite temperatures~\cite{mer:vla:pau:vie:17, mer:vla:pau:vie:17:qw}. On the contrary, the interferometric LE exhibits critical behaviour even at finite temperatures (confirming previous studies on DPTs~\cite{hey:bud:17, bah:ban:dut:17}).

\emph{Note added}. During the preparation of this manuscript, we became aware of a recent related work~\citep{sed:flei:sir:17}.

\acknowledgements{B. M. and C. V. acknowledge the support from DP-PMI and FCT (Portugal) through the grants SFRH/BD/52244/2013 and PD/BD/52652/2014, respectively. N. P. acknowledges the IT project QbigD funded by FCT PEst-OE/EEI/LA0008/2013, UID/EEA/50008/2013 and the Confident project  PTDC/EEI-CTP/4503/2014. Support from FCT (Portugal) through Grant UID/CTM/04540/2013 is also acknowledged. O.V. thanks Fundaci\'on Rafael del Pino, Fundaci\'on Ram\'on Areces and RCC Harvard.}
\\

\section{Appendix: Analytical Derivation of the Dynamical Susceptibilities} 
\label{sec:Los.vs.Fid}

\subsection{Zero Temperature case}

Let $\mathscr{H}$ be a Hilbert space. Suppose we have a family of Hamiltonians $\{H(\lambda):\lambda \in M\}$ where $M$ is a smooth  compact manifold of the Hamiltonian's parameters. We assume that aside from a closed finite subset of $M$, $C=\{\lambda_{i}\}_{i=1}^{n}\subset M$, the Hamiltonian is gapped and the ground state subspace is one-dimensional. Locally, on $M-C$, we can find a ground state (with unit norm) described by $\ket{\psi(\lambda)}$. Take $\lambda_i\in C$, and let $U$ be an open neighbourhood containing $\lambda_i$. Of course, for sufficiently small $U$, on the open set $U-\{\lambda_i\}$ one can find a smooth assignment $\lambda\mapsto \ket{\psi(\lambda)}$. Consider a curve $[0,1]\ni s\mapsto \lambda(s)\in U$, with initial condition $\lambda(0)=\lambda_0$, such that $\lambda(s_0)=\lambda_i$ for some $s_0\in[0,1]$. The family of Hamiltonians $H(s):=H(\lambda(s))$ is well-defined for every $s\in [0,1]$. The family of states $\ket{\psi(s)}\equiv \ket{\psi(\lambda(s))}$ is well-defined for $s\neq s_0$ and so is the ground state energy
\begin{align*}
E(s):=\bra{\psi(s)}H(s)\ket{\psi(s)}.
\end{align*} 
The overlap,
\begin{align*}
\mathcal{A}(s):=\bra{\psi(0)}\exp(-itH(s))\ket{\psi(0)},
\end{align*}
is well-defined. We can write
\begin{align*}
\exp(-itH(s))=\exp(-itH(0)) T\exp\left\{-i\int_{0}^{t} d\tau V(s,\tau)\right\}.
\end{align*}
If we take a derivative with respect to $t$ of the equation, we find,
\begin{align*}
H(s)=H(0) +\exp(-itH(0))V(s,t)\exp(itH(0))
\end{align*}
so,
\begin{align*}
V(s,t)=\exp(itH(0)) (H(s)-H(0))\exp(-itH(0)).
\end{align*}
We can now write, since $\ket{\psi(0)}$ is an eigenvector of $H(0)$,
\begin{align*}
\mathcal{A}(s)=e^{-it E(0)} \bra{\psi(0)}T\exp\left\{-i\int_{0}^{t} d\tau V(s,\tau)\right\}\ket{\psi(0)}.
\end{align*}
We now perform an expansion of the overlap
\begin{align*}
\bra{\psi(0)}T\exp\left\{-i\int_{0}^{t} d\tau V(s,\tau)\right\}\ket{\psi(0)}
\end{align*}
in powers of $s$. Notice that
\begin{align*}
 T \exp\left\{-i\int_{0}^{t} d\tau V(s,\tau)\right\}= I -i\int_{0}^{t} d\tau V(s,\tau)
 -\frac{1}{2}\int_{0}^{t}\int_{0}^{t}d\tau_2 d\tau_1 T\{V(s,\tau_2) V(s,\tau_1)\}+...
\end{align*}
and hence
\begin{align*}
\frac{d}{ds}\left(T \exp\left\{-i\int_{0}^{t} d\tau V(s,\tau)\right\}\right)\bigg|_{s=0}=-i\int_{0}^{t} d\tau \frac{\partial V}{\partial s}(0,\tau)
\end{align*}
and
\begin{align*}
\frac{d^2}{ds^2}\left(T \exp\left\{-i\int_{0}^{t} d\tau V(s,\tau)\right\}\right)\bigg|_{s=0}=-i\int_{0}^{t}d\tau \frac{\partial^2V}{\partial s^2}(0,\tau)-\int_{0}^{t}\int_{0}^t d\tau_2d\tau_1T\left\{\frac{\partial V}{\partial s}(0,\tau_2)\frac{\partial V}{\partial s}(0,\tau_1)\right\}.
\end{align*}
Therefore,
\begin{widetext}
\begin{small}
\begin{align*}
&\bra{\psi(0)}T\exp\left\{-i\int_{0}^{t} d\tau V(s,\tau)\right\} \ket{\psi(0)} = 
1-is\bra{\psi(0)}\int_{0}^{t} d\tau \frac{\partial V}{\partial s}(0,\tau)\ket{\psi(0)}\\
&+\frac{s^2}{2}\left[-i\bra{\psi(0)}\!\!\int_{0}^{t}\!\!\!d\tau \frac{\partial^2V}{\partial s^2}(0,\tau)\ket{\psi(0)}\! - \!\bra{\psi(0)}\!\!\int_{0}^{t}\!\!\!\int_{0}^t \!\!\! d\tau_2d\tau_1T\!\left\{\frac{\partial V}{\partial s}(0,\tau_2)\frac{\partial V}{\partial s}(0,\tau_1)\right\}\ket{\psi(0)}\right]
+\text{O}(s^3).
\end{align*}
\end{small}
\end{widetext}
Thus, by using the identity $\theta(\tau)+\theta(-\tau)=1$ of the Heaviside theta function, we obtain
\begin{widetext}
\begin{small}\begin{align*}
|\mathcal{A}(s)|^2=1-s^2\left[\int_{0}^{t}\int_{0}^{t}d\tau_2d\tau_1\bra{\psi(0)}\frac{1}{2}\left\{\frac{\partial V}{\partial s}(0,\tau_2),\frac{\partial V}{\partial s}(0,\tau_1)\right\}\ket{\psi(0)}
-\bra{\psi(0)}\frac{\partial V}{\partial s}(0,\tau_2)\ket{\psi(0)}\bra{\psi(0)}\frac{\partial V}{\partial s}(0,\tau_1)\ket{\psi(0)}\right]
+\text{O}(s^3).
\end{align*}\end{small}\end{widetext}
If we denote the expectation value $\bra{\psi(0)}\ast\ket{\psi(0)}\equiv\langle \ast\rangle$ we can write,
\begin{widetext}\begin{align*}
|\mathcal{A}(s)|^2=1-s^2\int_{0}^t\int_0^td\tau_2d\tau_1 \left[\langle\frac{1}{2}\left\{\frac{\partial V}{\partial s}(0,\tau_2),\frac{\partial V}{\partial s}(0,\tau_1)\right\}\rangle -\langle\frac{\partial V}{\partial s}(0,\tau_2)\rangle\langle\frac{\partial V}{\partial s}(0,\tau_1)\rangle \right]+\text{O}(s^3) = 1 - \chi s^2 + \text{O}(s^3),\end{align*}
where
\begin{align*}
\chi\equiv\int_{0}^t\int_0^td\tau_2d\tau_1 \left[\langle\frac{1}{2}\left\{\frac{\partial V}{\partial s}(0,\tau_2),\frac{\partial V}{\partial s}(0,\tau_1)\right\}\rangle -\langle\frac{\partial V}{\partial s}(0,\tau_2)\rangle\langle\frac{\partial V}{\partial s}(0,\tau_1)\rangle \right]
\end{align*}
\end{widetext}
is the dynamical susceptibility and is naturally nonnegative. In fact, defining $V_{a}(\tau)=e^{i\tau H(0)}\partial H/\partial \lambda^{a}(\lambda_0)e^{-i\tau H(0)}$ such that, by the chain rule,
\begin{align*}
\frac{\partial V}{\partial s}(0,\tau)=V_{a}(\tau)\frac{\partial\lambda^{a}}{\partial s}(0),
\end{align*}
we can write,
\begin{align*}
\chi=g_{ab}(\lambda_0)\frac{\partial\lambda^{a}}{\partial s}(0)\frac{\partial\lambda^{b}}{\partial s}(0),
\end{align*}
with the metric tensor given by
\begin{widetext}\begin{align}
g_{ab}(\lambda_0)=\int_{0}^t\int_0^td\tau_2d\tau_1 \left[\langle\frac{1}{2}\left\{V_a(\tau_2),V_b(\tau_1)\right\}\rangle -\langle V_a(\tau_2)\rangle\langle V_b(\tau_1)\rangle \right].
\label{eq:dis-metric}
\end{align}
\end{widetext}

\subsection{Dynamical fidelity susceptibility $\chi$ at finite temperature}

A possible generalisation of the zero-temperature LE to finite temperatures is through  the Uhlmann fidelity, since the zero temperature $|\mathcal{A}(s)|$ is precisely the fidelity between the states $\ket{\psi(0)}$ and $\exp(-itH(s))\ket{\psi(0)}$. Since the Uhlmann fidelity between two close mixed states is determined by the Bures metric, we begin by revisiting the derivation of the latter for the case of interest, i.e., two-level systems.

\subsection{Bures metric for a two-level system}

Take a curve of full-rank density operators $t\mapsto \rho(t)$ and an horizontal lift $t\mapsto W(t)$, with $W(0)=\sqrt{\rho(0)}$. Then the Bures metric is given by
\begin{align*}
g_{\rho(t)}(\frac{d\rho}{dt},\frac{d\rho}{dt})=\tr \left\{\frac{dW^{\dagger}}{dt}\frac{dW}{dt}\right\}.
\end{align*}
The horizontality condition is given by
\begin{align*}
W^{\dagger}\frac{dW}{dt}=\frac{dW}{dt}^{\dagger}W,
\end{align*}
for each $t$. In the full-rank case, we can find a unique Hermitian matrix $G(t)$, such that 
\begin{align*}
\frac{dW}{dt}=G(t)W
\end{align*}
solves the horizontality condition
\begin{align*}
W^{\dagger}\frac{dW}{dt}=W^{\dagger}GW=\frac{dW}{dt}^{\dagger}W.
\end{align*} 
Also, $G$ is such that
\begin{align*}
\frac{d\rho}{dt}=\frac{d}{dt}(WW^{\dagger})=G\rho+\rho G.
\end{align*}
If $L_{\rho}$ ($R_\rho$) is left  (right) multiplication by $\rho$, we have, formally,
\begin{align*}
G=(L_{\rho}+R_{\rho})^{-1}\frac{d\rho}{dt}.
\end{align*}
Therefore,
\begin{align*}
g_{\rho(t)}(\frac{d\rho}{dt},\frac{d\rho}{dt})=\tr \left\{\frac{dW^{\dagger}}{dt}\frac{dW}{dt}\right\}=\tr\left\{G^2\rho\right\}
=\frac{1}{2}\tr\{G(\rho G+G\rho)\}
=\frac{1}{2}\tr\{G\frac{d\rho}{dt}\}=\frac{1}{2}\tr\left\{(L_{\rho}+R_{\rho})^{-1}\frac{d\rho}{dt}\frac{d\rho}{dt}\right\}.
\end{align*}
If we write $\rho(t)$ in the diagonal basis,
\begin{align*}
\rho(t)=\sum_{i}p_i(t)\ket{i(t)}\bra{i(t)}
\end{align*}
we find
\begin{align*}
g_{\rho(t)}(\frac{d\rho}{dt},\frac{d\rho}{dt})=\frac{1}{2}\tr\left\{(L_{\rho}+R_{\rho})^{-1}\frac{d\rho}{dt}\frac{d\rho}{dt}\right\}
=\frac{1}{2}\sum_{i,j}\frac{1}{p_i(t)+p_j(t)}\bra{i(t)}\frac{d\rho}{dt}\ket{j(t)}\bra{j(t)}\frac{d\rho}{dt}\ket{i(t)}.
\end{align*}
Hence, using the diagonal basis of $\rho$, we can read off the metric tensor at $\rho$ as
\begin{align*}
g_{\rho}=\frac{1}{2}\sum_{i,j}\frac{1}{p_i+p_j}\bra{i}d\rho\ket{j}\bra{j}d\rho\ket{i}.
\end{align*}
This is the result for general full rank density operators. For two-level systems, writing
\begin{align*}
\rho=\frac{1}{2}(1-X^{\mu}\sigma_{\mu}),
\end{align*}
and defining variables $|X|=r$ and $n^{\mu}=X^{\mu}/|X|$, we can express $g_{\rho}$ as
\begin{align*}
g_{\rho}=\left[\frac{1}{1+r}+\frac{1}{1-r}\right]d\rho_{11}^2 +d\rho_{12}d\rho_{21}=\frac{1}{1-r^2}d\rho_{11}^2+d\rho_{12}d\rho_{21},
\end{align*}
where we used $d\rho_{11}=-d\rho_{22}$. Notice that
\begin{align*}
d\rho_{11}=\frac{1}{2}\tr\{d\rho U\sigma_{3}U^{-1}\}=\frac{1}{2}\tr\{d\rho n^{\mu}\sigma_{\mu}\},
\end{align*}
where $U$ is a unitary matrix diagonalizing $\rho$, $U\sigma_3U^{-1}=n^{\mu}\sigma_{\mu}$. Now,
\begin{align*}
d\rho=-\frac{1}{2}dX^{\mu}\sigma_{\mu},
\end{align*}
and hence
\begin{align*}
d\rho_{11}=-\frac{1}{2}dX^{\mu}n_{\mu}=-\frac{1}{2}dr.
\end{align*}
On the other hand,\begin{widetext}
\begin{align*}
d\rho_{12}d\rho_{21}&=\frac{1}{4}\tr\{d\rho U(\sigma_1-i\sigma_2)U^{-1}\}\tr\{d\rho U(\sigma_1-i\sigma_2)U^{-1}\}\\
&=\frac{1}{4}\left[\tr\{d\rho U\sigma_1U^{-1}\}\tr\{d\rho U\sigma_1U^{-1}\}+\tr\{d\rho U\sigma_2U^{-1}\}\tr\{d\rho U\sigma_2U^{-1}\}\right]\\
&=\frac{1}{4}\delta_{\mu\nu}(dX^{\mu}-n^{\mu}n_{\lambda}dX^{\lambda})(dX^{\nu}-n^{\nu}n_{\sigma}dX^{\sigma})=\frac{1}{4}r^2dn^{\mu}dn_{\mu},
\end{align*}\end{widetext}
where we used the fact that the vectors $(u,v)$ defined by the equations $U\sigma_1U^{-1}=u^{\mu}\sigma_{\mu}$ and $U\sigma_2U^{-1}=v^{\mu}\sigma_{\mu}$ form an orthonormal basis for the orthogonal complement in $\mathbb{R}^3$ of the line generated by $n^{\mu}$ (which corresponds to the tangent space to the unit sphere $S^2$ at $n^{\mu}$). Thus, we obtain the final expression for the squared line element
\begin{align}
ds^2=\frac{1}{4}\left(\frac{dr^2}{1-r^2}+r^2\delta_{\mu\nu}dn^{\mu}dn^{\nu}\right).
\label{Eq: Bures Metric}
\end{align}

The above expression is ill-defined for the pure state case of $r=1$. Nevertheless, the limiting case of $r\to 1$ as the metric is smooth as we will now show by introducing another coordinate patch. The set of pure states is defined by $r=1$, i.e., they correspond to the boundary of the $3$-dimensional ball $B=\{X \ :\  |X| = r \leq 1\}$ which, topologically, is the set of all density matrices in dimension $2$. Introducing the change of variable $r = \cos u$, with $u \in [0,\pi /2)$, the metric becomes 
\begin{align*}
ds^2=\frac{1}{4}\left(du^2 + (\cos u)^2\delta_{\mu\nu}dn^{\mu}dn^{\nu}\right),
\end{align*}
which is well defined also for the pure-state case of $r=\cos(0) = 1$. Restricting it to the unit sphere, the metric coincides with the Fubini-Study metric, also known as the quantum metric, on the space of pure states $\mathbb{C}P^1\cong S^2$, i.e., the Bloch sphere. Therefore, there is no problem on taking the pure-state limit of this metric on the space of states, since it reproduces the correct result.

\subsection{Pullback of the Bures metric}

We have a map
\begin{align*}
M\ni \lambda\mapsto \rho(\lambda)=U(\lambda)\rho_0U(\lambda)^{-1}=\frac{1}{2}U(\lambda)\left(I -X^\mu\sigma_\mu\right)U(\lambda)^{-1},
\end{align*}
with
\begin{align*}
U(\lambda)=\exp(-itH(\lambda)),
\end{align*}
and we take
\begin{align*}
\rho_0=\frac{\exp(-\beta H(\lambda_{0}))}{\tr\left\{\exp(-\beta H(\lambda_{0})\right\}}, \text{ for some }\lambda_0\in M.
\end{align*}
We use the curve $[0,1] \ni s\mapsto \lambda(s)$, with $\lambda(0)=\lambda_0$, to obtain a curve of density operators
\begin{align*}
s\mapsto \rho(s):=\rho(\lambda(s)).
\end{align*}
Notice that $\rho(0)=\rho_0$. The fidelity we consider is then
\begin{align*}
F(\rho(0),\rho(s))=\tr\left(\sqrt{\sqrt{\rho(0)}\rho(s)\sqrt{\rho(0)}}\right).	
\end{align*}
Recall that for $2\times 2$ density operators of full rank the Bures line element reads
\begin{align*}
ds^2=\frac{1}{4}\left(\frac{dr^2}{1-r^2}+ r^2 \delta_{\mu\nu} dn^{\mu}dn^{\nu}\right),
\end{align*}
where
\begin{align*}
n^{\mu}=X^{\mu}/|X| \text{ and } r=|X|,
\end{align*}
with
\begin{align*}
\rho= \frac{1}{2}\left(I-X^{\mu}\sigma_\mu\right).
\end{align*}
Now,
\begin{align*}
\rho(\lambda)=\frac{1}{2}\left(I- R^{\mu}_{\ \nu}(\lambda) X^{\nu}\sigma_{\mu}\right),
\end{align*}
with $R^{\mu}_{\ \nu}(\lambda)$ being the unique $\text{SO}(3)$ element satisfying
\begin{align*}
U(\lambda)\sigma_{\mu}U(\lambda)^{-1}=R_{\ \mu}^{\nu}(\lambda)\sigma_{\nu}.
\end{align*}
We then have, pulling back the coordinates,
\begin{align*}
r(\lambda)=|X|=\text{constant} \  \text{ and } \  n^{\mu}(\lambda)=R^{\mu}_{\ \nu}(\lambda)n^{\nu}. \end{align*}
Therefore,
\begin{align*}
ds^2=\frac{1}{4}r^2\delta_{\mu\nu}\frac{\partial n^{\mu}}{\partial \lambda^{a}}\frac{\partial n^{\nu}}{\partial \lambda^{b}}d\lambda^{a}d\lambda^{b}=\frac{1}{4}r^2\delta_{\mu\nu}n^{\sigma}n^{\tau}\frac{\partial R^{\mu}_{\ \sigma}}{\partial \lambda^{a}}\frac{\partial R^{\nu}_{\ \tau}}{\partial \lambda^{b}}d\lambda^{a}d\lambda^{b},
\end{align*}
which in terms of the Euclidean metric on the tangent bundle of $\mathbb{R}^3$, denoted by $\langle \ast , \ast\rangle$, takes the form
\begin{align*} 
g_{ab}(\lambda)=\frac{1}{4} r^2\langle R^{-1}\frac{\partial R}{\partial \lambda^{a}}n,  R^{-1}\frac{\partial R}{\partial \lambda^{b}} n\rangle,
\end{align*}
written in terms of the pullback of the Maurer-Cartan form in $\text{SO}(3)$, $R^{-1}dR$. We can further pullback by the curve $s\mapsto \lambda(s)$ and evaluate at $s=0$
\begin{align*}
\chi:=g_{ab}(\lambda_0)\frac{\partial \lambda^a}{\partial s}(0)\frac{\partial \lambda^{b}}{\partial s}(0)=\frac{1}{4} r^2\langle R^{-1}\frac{\partial R}{\partial \lambda^{a}}n,  R^{-1}\frac{\partial R}{\partial \lambda^{b}} n\rangle\frac{\partial \lambda^a}{\partial s}(0)\frac{\partial \lambda^{b}}{\partial s}(0),
\end{align*}
which gives us the expansion of the fidelity
\begin{align*}
F(s)\equiv F(\rho(0),\rho(s))=1-\frac{1}{2}\chi s^2 +...
\end{align*}
We now evaluate $\chi$. Note that
\begin{align*}
dU\sigma_{\mu}U^{-1} +U\sigma_{\mu}dU^{-1} = U[U^{-1}dU,\sigma_{\mu}]U^{-1} = dR^{\nu}_{\ \mu}\sigma_\nu.
\end{align*}
Now, we can parameterise
\begin{align*}
U=y^{0}I+iy^{\mu}\sigma_{\mu}, \text{ with } |y|^2=1. 
\end{align*}
Therefore,
\begin{align*}
U^{-1}dU=(y^0 -iy^{\mu}\sigma_\mu)(dy^0 +idy^{\nu}\sigma_{\nu})
=i(y^0dy^{\mu}-y^{\mu}dy^{0})\sigma_{\mu}+\frac{i}{2}(y^{\mu}dy^{\nu} - y^{\nu}dy^{\mu})\varepsilon_{\mu\nu}^{\lambda}\sigma_{\lambda};\end{align*},
\begin{widetext}\begin{align*}
[U^{-1}dU,\sigma_{\kappa}]&=-2\left[(y^0dy^{\mu}-y^{\mu}dy^{0})\varepsilon_{\mu\kappa}^{\tau}+\frac{1}{2}(y^{\mu}dy^{\nu} - y^{\nu}dy^{\mu})\varepsilon_{\mu\nu}^{\lambda}\varepsilon_{\lambda\kappa}^{\tau}\right]\sigma_{\tau}\\
&=-2\left[(y^0dy^{\mu}-y^{\mu}dy^{0})\varepsilon_{\mu\kappa}^{\ \ \ \tau}+\frac{1}{2}(y^{\mu}dy^{\nu} - y^{\nu}dy^{\mu})(\delta_{\mu\kappa}\delta^{\tau}_{\nu}-\delta_{\mu}^{\tau}\delta^{\nu}_{\kappa})\right]\sigma_{\tau}\\
&=-2\left[(y^0dy^{\mu}-y^{\mu}dy^{0})\varepsilon_{\mu\kappa}^{\ \ \ \tau}+(y^{\kappa}dy^{\tau}-y^{\tau}dy^{\kappa})\right]\sigma_{\tau}\\
&=2\left[(y^0dy^{\mu}-y^{\mu}dy^{0})\varepsilon_{\mu\ \ \kappa}^{\ \tau}+(y^{\tau}dy^{\kappa}-y^{\kappa}dy^{\tau})\right]\sigma_{\tau}\equiv (R^{-1}dR)^{\tau}_{\ \kappa}\sigma_{\tau}.
\end{align*}\end{widetext}
Observe that for $H(\lambda)=x^{\mu}(\lambda)\sigma_{\mu}$ we have,
\begin{align*}
y^0(\lambda)=\cos(|x(\lambda)|t) \text{ and } y^{\mu}=-\sin(|x(\lambda)|t)\frac{x^{\mu}(\lambda)}{|x(\lambda)|}.
\end{align*}
Therefore,
\begin{align*}
dy^{0} & =-\sin(|x(\lambda)|t)d|x(\lambda)|,\\
dy^{\mu} & =-\cos(|x(\lambda)|t)\frac{x^{\mu}(\lambda)}{|x(\lambda)|}d|x(\lambda)|-\sin(|x(\lambda)|t)d\left(\frac{x^{\mu}(\lambda)}{|x(\lambda)|}\right).
\end{align*}
After a bit of algebra, we get
\begin{align*}
y^0dy^{\mu}-y^{\mu}dy^0  = &\ \frac{x^{\mu}(\lambda)}{|x(\lambda)|}d|x(\lambda)|-\sin(|x(\lambda)|)\cos(|x(\lambda)|)d\left(\frac{x^{\mu}(\lambda)}{|x(\lambda)|}\right),
\\
y^{\mu}dy^{\nu}-y^{\nu}dy^{\mu} = & \ 2\sin^2(|x(\lambda)|t)\frac{x^{[\mu}(\lambda)}{|x(\lambda)|}d\left(\frac{x^{\nu]}(\lambda)}{|x(\lambda)|}\right).
\end{align*}
Thus,
\begin{align*}
(R^{-1}dR)^{\tau}_{\kappa}=2\frac{x^{\mu}(\lambda)}{|x(\lambda)|}d|x(\lambda)|\varepsilon_{\mu\kappa}^{\ \tau}-\sin(2|x(\lambda)|t)d\left(\frac{x^{\mu}(\lambda)}{|x(\lambda)|}\right)\varepsilon_{\mu\kappa}^{\ \tau}+4\sin^2(|x(\lambda)|t)\frac{x^{[\tau}(\lambda)}{|x(\lambda)|}d\left(\frac{x^{\kappa]}(\lambda)}{|x(\lambda)|}\right).
\end{align*}
At $s=0$, $\lambda(0)=\lambda_0$ and the coordinate $n^{\mu}(\lambda(0))=x^{\mu}(\lambda_0)/|x(\lambda_0)|$, so the previous expression reduces to
\begin{align*}
(R^{-1}dR(\lambda_0))^{\tau}_{\ \kappa}n^{\kappa}=(R^{-1}dR(\lambda_0))^{\tau}_{\ \kappa}\frac{x^{\kappa}(\lambda_0)}{|x(\lambda_0)|}
=  -\sin(2|x(\lambda_0)|t)\frac{1}{|x(\lambda_0)|^2}\varepsilon^{\tau}_{\ \ \mu \kappa}x^{\mu}(\lambda_0)dx^{\kappa}(\lambda_0)- (1-\cos(2|x(\lambda_0)|t))d\left(\frac{x^{\mu}}{|x|}\right)(\lambda_0).
\end{align*}
Notice that the first term is perpendicular to the second. Therefore, we find
\begin{widetext}\begin{align*}
\chi ds^2 &=\frac{1}{4}r^2|R^{-1}dR n|^2 \\ & =\frac{1}{4}r^2\left[\sin^2(2|x(\lambda_0)|t)\frac{1}{|x(\lambda_0)|^4}\left(\delta^{\lambda}_{\mu}\delta^{\sigma}_{\kappa}-\delta^{\sigma}_{\mu}\delta^{\lambda}_{\kappa}\right)x^{\mu}(\lambda)dx^{k}(\lambda)x_{\lambda}(\lambda_0)dx_{\sigma}(\lambda_0)+\left(1-\cos(2|x(\lambda_0|t)^2\langle P dx(\lambda_0),Pdx(\lambda_0)\rangle\right)\right]\\
&=r^2\frac{\sin^2(|x(\lambda_0)|t)}{|x(\lambda_0)|^2}\langle P \frac{\partial x}{\partial \lambda^a}(\lambda_0),P\frac{\partial x}{\partial \lambda^b}(\lambda_0)\rangle\frac{\partial \lambda^a}{\partial s}(0)\frac{\partial \lambda^b}{\partial s}(0)ds^2,
\end{align*}\end{widetext}
where we have introduced the projector $P:T_{x}\mathbb{R}^3=T_{x}S^{2}_{|x|}\oplus N_{x}S^{2}_{|x|}\to T_{x}S^{2}_{|x|}$ onto the tangent space of the sphere of radius $|x|$ at $x$. In other words, the pullback metric by $\rho$ of the Bures metric at $\lambda_0$ is given by
\begin{align*}
g_{ab}(\lambda_0)=r^2\frac{\sin^2(|x(\lambda_0)|t)}{|x(\lambda_0)|^2}\langle P \frac{\partial x}{\partial \lambda^a}(\lambda_0),P\frac{\partial x}{\partial \lambda^b}(\lambda_0)\rangle=\tanh^2(\beta |x(\lambda_0)|)\frac{\sin^2(|x(\lambda_0)|t)}{|x(\lambda_0)|^2}\langle P \frac{\partial x}{\partial \lambda^a}(\lambda_0),P\frac{\partial x}{\partial \lambda^b}(\lambda_0)\rangle.
\end{align*}

\subsection{Dynamical interferometric susceptibility $\tilde{\chi}$ at finite temperature}

We can replace the average $\langle e^{-itH(s)}\rangle\equiv \bra{\psi(0)}e^{-itH(s)}\ket{\psi(0)}$ by the corresponding average of $e^{itH(0)}e^{-itH(s)}$ on the mixed state $\rho(\lambda_0)=\rho(0)=\exp(-\beta H(0))/\tr\{\exp(-\beta H(0))\}$ (note its implicit temperature dependence):
\begin{align*}
\mathcal{A}(s)=\tr\left\{\rho(0)T\exp\left\{-i\int_{0}^{t} d\tau V(s,\tau)\right\}\right\}.
\end{align*}
It is easy to see that $|\mathcal{A}(s)|^2$ has the same expansion as before with the average on $\ket{\psi(0)}$ replaced by the average on $\rho(0)$.

We now proceed to compute $\tilde{\chi}$, or equivalently $\tilde{g}_{ab}(\lambda_0)$, in the case of a two-level system, where we can write
\begin{align*}
\rho(\lambda)=\frac{e^{-\beta H(\lambda)}}{\tr\{e^{-\beta H(\lambda)}\}}=\frac{1}{2}(I-X^{\mu}(\lambda)\sigma_{\mu}),
\end{align*}
and define variables $r(\lambda)=|X(\lambda)|$ and $n^{\mu}(\lambda)=X^{\mu}(\lambda)/|X(\lambda)|$.
Writing $H(\lambda)=x^{\mu}(\lambda)\sigma_{\mu}$ (and $H(s)\equiv H(\lambda(s))$), we have
\begin{align*}
V_{a}(\tau)=\frac{\partial x^{\mu}}{\partial \lambda^{a}}(\lambda_0)e^{i\tau H(0)}\sigma_{\mu} e^ {-i\tau H(0)}.
\end{align*}
Hence, its expectation value is
\begin{align*}
\langle V_{a}(\tau)\rangle =\frac{1}{\tr\{e^{-\beta H(\lambda)}\}} \tr\left\{e^{-\beta H(0)}\sigma_{\mu}\right\}\frac{\partial x^{\mu}}{\partial \lambda^{a}}(\lambda_0)=r(\lambda_0)n_{\mu}(\lambda_0)\frac{\partial x^{\mu}}{\partial \lambda^{a}}(\lambda_0)=X_{\mu}(\lambda_0)\frac{\partial x^{\mu}}{\partial \lambda^a}(\lambda_0),
\end{align*}
which is independent of $\tau$. We then have
\begin{align*}
&\langle V_{a}(\tau_2)\rangle \langle V_{b}(\tau_1)\rangle =(r(\lambda_0))^2n_{\mu}(\lambda_0)\frac{\partial x^{\mu}}{\partial \lambda^{a}}(\lambda_0)n_{\nu}(\lambda_0)\frac{\partial x^{\nu}}{\partial \lambda^{b}}(\lambda_0)\\
&=\tanh^2(\beta|x(\lambda_0)|)\frac{x_{\mu}}{|x(\lambda_0)|}\frac{\partial x^{\mu}}{\partial \lambda^{a}}(\lambda_0)\frac{x_{\nu}}{|x(\lambda_0)|}\frac{\partial x^{\nu}}{\partial \lambda^{b}}(\lambda_0),
\end{align*}
where we used $X^{\mu}(\lambda_0)=\tanh(\beta |x(\lambda_0)|)x^{\mu}(\lambda_0)/|x(\lambda_0)|$.
Now, using the cyclic property of the trace, we get
\begin{widetext}\begin{small}
\begin{align}
\!\!\!\!\!\!\!\!\frac{1}{2\tr\{e^{-\beta H(0)}\}}\tr \left\{ e^{-\beta H(\lambda)}\{V_{a}(\tau_2),V_{b}(\tau_1)\} \right\}&=
\frac{1}{2\tr\{e^{-\beta H(0)}\}}\tr \left\{ e^{-\beta H(0)}\{\sigma_{\mu},\sigma_{\nu}\}\right\}R^{\mu}_{\ \lambda}(\tau_2)R^{\nu}_{\ \sigma}(\tau_1)\frac{\partial x^{\lambda}}{\partial \lambda^{a}}(\lambda_0)\frac{\partial x^{\sigma}}{\partial \lambda^{b}}(\lambda_0)\nonumber\\
&=\delta_{\mu\nu}R^{\mu}_{\ \lambda}(\tau_2)R^{\nu}_{\ \sigma}(\tau_1)\frac{\partial x^{\lambda}}{\partial \lambda^{a}}(\lambda_0)\frac{\partial x^{\sigma}}{\partial \lambda^{b}}(\lambda_0),
\label{eq:anti-com}
\end{align}
\end{small}\end{widetext}
where $R^{\mu}_{\ \nu}(\tau)$ is the rotation matrix defined by the equation
\begin{align}
e^{i\tau H(0)}\sigma_{\nu}e^{-i\tau H(0)}=R^{\mu}_{\ \nu}(\tau)\sigma_{\mu}.
\label{eq:su(2)toso(3)}
\end{align}
We can explicitly write $R^{\mu}_{\ \nu}(\tau)$ as
\begin{align*}
R^{\mu}_{\nu}(\tau)=\cos(2\tau |x(\lambda_0)|)\delta^{\mu}_{\nu} +(1-\cos(2\tau|x(\lambda_0)|))n^{\mu}(\lambda_0)n_{\nu}(\lambda_0)+\sin(2\tau |x(\lambda_0)|)n^{\lambda}(\lambda_0)\varepsilon_{\lambda\nu}^{\mu}.
\end{align*}
Using the previous equation, and because $\{R(\tau)\}$ forms a one-parameter group, we can write
\begin{align*}
\delta_{\mu\nu}R^{\mu}_{\ \lambda}(\tau_2)R^{\nu}_{\ \sigma}(\tau_1)=\delta_{\kappa\lambda}R^{\kappa}_{\ \sigma }(\tau_2-\tau_1).
\end{align*}
Since  $\tilde{\chi}$ (i.e., the metric $\tilde{g}_{ab}$; recall its zero-temperature expression~\eqref{eq:dis-metric}) has to be symmetric under the label exchange $a\leftrightarrow b$, the relevant symmetric part of~\eqref{eq:anti-com} is
\begin{small}\begin{align*}
\cos\left[2(\tau_2-\tau_1)|x(\lambda_0)|\right]\delta_{\mu\nu}\frac{\partial x^{\mu}}{\partial \lambda^{a}}(\lambda_0)\frac{\partial x^{\nu}}{\partial \lambda^{b}}(\lambda_0)
+(1-\cos\left[2(\tau_2-\tau_1)|x(\lambda_0)|\right])\frac{x_{\mu}}{|x(\lambda_0)|}\frac{\partial x^{\mu}}{\partial \lambda^{a}}(\lambda_0)\frac{x_{\nu}}{|x(\lambda_0)|}\frac{\partial x^{\nu}}{\partial \lambda^{b}}(\lambda_0).
\end{align*}\end{small}
Putting everything together gives
\begin{widetext}\begin{align*}
\langle\frac{1}{2}\left\{V_a(\tau_2),V_b(\tau_1)\right\}\rangle -\langle V_a(\tau_2)\rangle\langle V_b(\tau_1)\rangle
=&\cos\left[2(\tau_2-\tau_1)|x(\lambda_0)|\right]\langle P\frac{\partial x}{\partial \lambda^a} (\lambda_0),P\frac{\partial x}{\partial \lambda^b} (\lambda_0)\rangle  \\
&+(1-\tanh^2(\beta|x(\lambda_0)|)) \langle \frac{x(\lambda_0)}{|x(\lambda_0)|},\frac{\partial x}{\partial \lambda^a}(\lambda_0) \rangle\langle \frac{x(\lambda_0)}{|x(\lambda_0)|},\frac{\partial x}{\partial \lambda^b}(\lambda_0)\rangle.
\end{align*}

The integral on $\tau_1$ and $\tau_2$ can now be performed, using
\begin{align*}
&\int_{0}^{t}\int_{0}^{t}d\tau_2d\tau_1\cos[2(\tau_2-\tau_1)\varepsilon]=\int_{0}^{t}\int_{0}^{t}d\tau_2d\tau_1\left(\cos(2\tau_2\epsilon)\cos(2\tau_1\epsilon)+\sin(2\tau_2\epsilon)\sin(2\tau_1\epsilon)\right)\\
&=\frac{1}{4\epsilon^2}\left[\sin^2(2t\epsilon) + (\cos(2t\epsilon)-1)(\cos(2t\epsilon)-1)\right]=\frac{1}{4\epsilon^2}\left[2-2\cos(2t\epsilon)\right]=\frac{\sin^2(t\epsilon)}{\epsilon^2}.
\end{align*}\end{widetext}
So, the interferometric metric is
\begin{align*}
\tilde{g}_{ab}(\lambda_0)=\frac{\sin^2(|x(\lambda_0)|t)}{|x(\lambda_0)|^2}[\langle P \frac{\partial x}{\partial \lambda^a}(\lambda_0),P\frac{\partial x}{\partial \lambda^b}(\lambda_0)\rangle]+t^2(1-\tanh^2(\beta |x(\lambda_0|))\langle \frac{x(\lambda_0)}{|x(\lambda_0)|},\frac{\partial x}{\partial \lambda^a}(\lambda_0) \rangle\langle \frac{x(\lambda_0)}{|x(\lambda_0)|},\frac{\partial x}{\partial \lambda^b}(\lambda_0)\rangle.
\end{align*}
The dynamical interferometric susceptibility is then given by
\begin{widetext}\begin{align*}
\tilde{\chi}&=\tilde{g}_{ab}(\lambda_0)\frac{\partial \lambda^a}{\partial s}(0)\frac{\partial \lambda^b}{\partial s}(0)=\int_{0}^t\int_0^td\tau_2d\tau_1 \left[\langle\frac{1}{2}\left\{\frac{\partial V}{\partial s}(0,\tau_2),\frac{\partial V}{\partial s}(0,\tau_1)\right\}\rangle -\langle\frac{\partial V}{\partial s}(0,\tau_2)\rangle\langle\frac{\partial V}{\partial s}(0,\tau_1)\rangle \right]\\
&=\{\frac{\sin^2(|x(\lambda_0)|t)}{|x(\lambda_0)|^2}[\langle P \frac{\partial x}{\partial \lambda^a}(\lambda_0),P\frac{\partial x}{\partial \lambda^b}(\lambda_0)\rangle]+t^2(1-\tanh^2(\beta |x(\lambda_0|))\langle \frac{x(\lambda_0)}{|x(\lambda_0)|},\frac{\partial x}{\partial \lambda^a}(\lambda_0) \rangle\langle \frac{x(\lambda_0)}{|x(\lambda_0)|},\frac{\partial x}{\partial \lambda^b}(\lambda_0)\rangle\}\frac{\partial \lambda^a}{\partial s}(0)\frac{\partial \lambda^b}{\partial s}(0),
\end{align*}\end{widetext}
with the average taken with respect to the thermal state $\rho_0=\rho(0)\equiv \rho(\lambda_0)$. Also, as mentioned previously, we have the expansion
\begin{align*}
|\mathcal{A}(s)|^2 =|\tr \left[\rho(0)\exp(it H(0))\exp(-it H(s))\right]|^2=\left|\tr\left[\rho(0) T\exp(-i\int_{0}^{t}d\tau V(s,\tau))\right]\right|^2=1-\tilde{\chi}s^2 +...
\end{align*}
The difference between the two susceptibilities is given by:
\begin{small}\begin{align*}
\tilde{\chi}-\chi &=  (1-\tanh^2(\beta|x(\lambda_0)|))\{\frac{\sin^2(|x(\lambda_0)|t)}{|x(\lambda_0)|^2}[\langle P \frac{\partial x}{\partial \lambda^a}(\lambda_0),P\frac{\partial x}{\partial \lambda^b}(\lambda_0)\rangle]\\
&+t^2\langle \frac{x(\lambda_0)}{|x(\lambda_0)|},\frac{\partial x}{\partial \lambda^a}(\lambda_0) \rangle\langle \frac{x(\lambda_0)}{|x(\lambda_0)|},\frac{\partial x}{\partial \lambda^b}(\lambda_0)\rangle\}\frac{\partial \lambda^a}{\partial s}(0)\frac{\partial \lambda^b}{\partial s}(0).
\end{align*}\end{small}
As $\beta\to +\infty$, i.e., as the temperature goes to zero, the two susceptibilities are equal. Now, the function
\begin{align*}
f(t)=\frac{\sin^2(\epsilon t)}{\epsilon^2}
\end{align*}
is well approximated by $t^2$ for small enough $\epsilon$. In that case the sum of the two terms appearing in the difference between susceptibilities is just proportional the pull-back Euclidean metric on $T\mathbb{R}^3$.

\subsection{The pullback of the inteferometric (Riemannian) metric on the space of unitaries}

We first observe that each full rank density operator $\rho$ defines a Hermitian inner product in the vector space of linear maps of a Hilbert space $\mathscr{H}$, i.e., $\text{End}(\mathscr{H})$, given by,
\begin{align*}
\langle A,B\rangle_{\rho}\equiv \tr \{\rho A^{\dagger}B\}.
\end{align*}
This inner product then defines a Riemannian metric on the trivial tangent bundle of the vector space $\text{End}(\mathscr{H})$. Since the unitary group $\text{U}(\mathscr{H})\subset \text{End}(\mathscr{H})$, by restriction we get a Riemannian metric on $\text{U}(\mathscr{H})$. If we choose $\rho$ to be $e^{-\beta H(\lambda)}/\tr \{e^{-\beta H(\lambda)}\}$, then take the pullback by the map $\Phi_{t}: M\ni \lambda_f\mapsto e^{-it H(\lambda_f)}\in\text{U}(\mathscr{H})$ and evaluate at $\lambda_f=\lambda$, to obtain the desired metric.

Next, we show that this version of LE is closely related to the interferometric geometric phase introduced by Sj\"{o}qvist \emph{ et. al}~\cite{sjo:pat:eke:ana:eri:oi:ved:00,ton:sjo:kwe:oh}. To see this, consider the family of distances in $\text{U}(\mathscr{H})$, $d_{\rho}$, parametrised by a full rank density operator $\rho$, defined as
\begin{align*}
d^2_{\rho}(U_1,U_2)=\tr\{\rho (U_1-U_2)^{\dagger}(U_1-U_2)\}=2(1-\text{Re}\langle U_1, U_2\rangle_{\rho}),
\end{align*}
where $\langle \ast , \ast \rangle_{\rho}$ is the Hermitian inner product defined previously. In terms of the spectral representation of $\rho=\sum_{j}p_j\ket{j}\bra{j}$, we have
\begin{align*}
\langle U_1, U_2\rangle_{\rho}=\sum_{j}p_j\bra{j} U_1^{\dagger} U_2 \ket{j}.
\end{align*}
The Hermitian inner product is invariant under $U_{i}\mapsto U_{i}\cdot D$, $i=1,2$, where $D$ is a phase matrix
\begin{align*}
D=e^{i\alpha}\sum_{j}\ket{j}\bra{j}.
\end{align*}
For the interferometric geometric phase, one enlarges this gauge symmetry to the subgroup of unitaries preserving $\rho$, i.e., the gauge degree of freedom is $U(1)\otimes\cdots\otimes U(1)$. However, since we are interested in the interferometric LE previously defined, we choose not to do that, as we only need the diagonal subgroup, i. e., we only have a global phase. Next, promoting this global $\text{U}(1)$-gauge degree of freedom to a local one, i.e., demanding that we only care about unitaries modulo a phase, we see that, upon changing $U_i\mapsto U_i\cdot D_i$, $i=1,2$,  we have
\begin{align*}
\langle U_1, U_2\rangle_{\rho}\mapsto & \langle U_1\cdot D_1, U_2\cdot D_2\rangle_{\rho}=\sum_{j}p_j\bra{j} U_1^{\dagger} U_2 \ket{j} e^{i(\alpha_{2}-\alpha_{1})}.
\end{align*}
We can choose gauges, i.e., $D_1$ and $D_2$, minimizing $d^2_{\rho}(U_1\cdot D_1,U_2\cdot D_2)$, obtaining
\begin{align*}
d^2_{\rho}(U_1\cdot D_1, U_2\cdot D_2)=2(1-|\langle U_1\cdot D_1,U_2\cdot D_2\rangle_\rho|)=2(1-|\langle U_1,U_2\rangle_\rho|).
\end{align*}
Now, if $\{U_i=U(t_i)\}_{1\leq i\leq N}$ were the discretisation of a path of unitaries $t\mapsto U(t)$, $t\in[0,1]$, applying the minimisation process locally, i.e., between adjacent unitaries $U_{i+1}$ and $U_i$, in the limit $N\to \infty$ we get a notion of parallel transport on the principal bundle $\text{U}(\mathscr{H})\to \text{U}(\mathscr{H})/\text{U}(1)$. In particular, the parallel transport condition reads as
\begin{align*}
\tr\left\{ \rho U^{\dagger}(t)\frac{dU}{dt}(t)\right\}=0, \text{ for all } t\in[0,1].
\end{align*}
If we take $\rho=\exp(-\beta H(\lambda_i))/\tr\{e^{-\beta H(\lambda_i)}\}$, $U_1=\exp(-it H(\lambda_i))$ and $U_2=\exp(-it H(\lambda_f))$, then the interferometric LE is
\begin{align*}
\mathcal{L}(t,\beta;\lambda_f,\lambda_i)=|\langle U_1,U_2\rangle_{\rho}|=\langle \widetilde{U}_1, \widetilde{U}_2\rangle_{\rho},
\end{align*}
where $\widetilde{U}_i=U_i\cdot D_i$ ($i=1,2$) correspond to representatives satisfying the discrete version of the parallel transport condition.

\bibliographystyle{unsrt}
\bibliography{mybib}
\end{document}